\documentclass[12pt]{article}

\usepackage{graphicx}  

\usepackage{amsmath}
\usepackage{amsfonts}

\renewcommand{\baselinestretch} {1.3}

\makeatletter
\def\singlespace{\def\baselinestretch{1}\@normalsize}

\newcommand{\thefont}[2]{\fontsize{#1}{#2}\fontshape{n}\selectfont}
\newcommand{\1}{\rlap{\thefont{10pt}{12pt}1}\kern.16em\rlap{\thefont{11pt}{13.2pt}1}\kern.4em}
\newtheorem{algorithm}{Algorithm}

\newcommand{\twofig}[4]
{ \hbox to\hsize{\hss
     \vbox{\psfig{figure=#1,width=#3,height=#4}}\qquad
     \vbox{\psfig{figure=#2,width=#3,height=#4}}
     \hss}
\vskip -0.0truein \hbox to\hsize{\hss
     \vbox{ \begin{center}\mbox{\footnotesize \hspace{0.0in} {(a)}
                      \hspace{#3} {(b)}  }  \end{center} }
     \hss}
\vskip 0.0truein }
\newcommand{\threefig}[5]{
\hbox to\hsize{\hss
     \vbox{\psfig{figure=#1,width=#4,height=#5}} \hspace{0.0in}
     \vbox{\psfig{figure=#2,width=#4,height=#5}} \hspace{0.2in}
     \vbox{\psfig{figure=#3,width=#4,height=#5}}
     \hss}
\vskip -0.0truein \hbox to\hsize{\hss
     \vbox{ \begin{center}\mbox{\footnotesize \hspace{0.0in} {(a)}
                      \hspace{#4} {(b)}   \hspace{#4} {(c)}}  \end{center} }
     \hss}
}
\newcommand{\fourfig}[6]
{ \hbox to\hsize{\hss
     \vbox{\psfig{figure=#1,width=#5,height=#6}}\qquad
     \vbox{\psfig{figure=#2,width=#5,height=#6}}
     \hss}
\vskip -0.0truein \hbox to\hsize{\hss
     \vbox{ \begin{center}\mbox{\footnotesize \hspace{0.1in} {(a)}
                      \hspace{#5} {(b)}  }  \end{center} }
     \hss}
\vskip 0.1truein \hbox to\hsize{\hss
     \vbox{\psfig{figure=#3,width=#5,height=#6}}\qquad
     \vbox{\psfig{figure=#4,width=#5,height=#6}}
     \hss}
\vskip -0.1truein
     \vbox{ \begin{center}\mbox{\footnotesize \hspace{0.1in} {(c)}
                      \hspace{#5} {(d)}  }  \end{center} }
\hbox to\hsize{\hss
     \hss}
\vskip -0.1truein }

\newcommand{\reals}{\ensuremath{{\mathbb R}}}
\newcommand{\RR}{\reals}

\newcommand{\PP}{\ensuremath{{\mathbb P}}}
\newcommand{\EE}{\ensuremath{{\mathbb E}}}

\newcommand{\btheta}{\mbox{$\boldsymbol{\theta}$}}

\newcommand{\bepsilon}{\mbox{\boldmath$\epsilon$}}

\newtheorem{theo}{Theorem}
\newtheorem{prop}{Proposition}
\newtheorem{lemma}{Lemma}

\begin{document}

\title{\sc Noisy Independent Factor Analysis Model for
Density Estimation and Classification}

\author{U.~Amato, A.~Antoniadis, A.~Samarov and A.B.~Tsybakov\thanks{U.~Amato, the Istituto per le Applicazioni del Calcolo
`M.~Picone' CNR, Napoli 80131, Italy (e-mail: u.amato@iac.cnr.it);
A.~Antoniadis, Laboratoire Jean Kuntzmann, Universit\'e Joseph
Fourier, Grenoble 38041, France (e-mail:
anestis.antoniadis@imag.fr); A. ~Samarov, Department of Mathematical
Sciences, University of Massachusetts Lowell and Sloan School of
Management, MIT, MA 02139 (e-mail: samarov@mit.edu); A.B.~Tsybakov,
Laboratoire de Statistique, CREST, Malakoff,  92240 France, and LPMA
(UMR CNRS 7599), Universit\'e Paris 6, Paris, France
 (e-mail:  alexandre.tsybakov@upmc.fr). Financial support from
the IAP research network of the Belgian government (Belgian Federal
Science Policy) is gratefully acknowledged. Research of A.~Samarov
was partially supported by NSF grant DMS-0505561 and by a grant from
Singapore-MIT Alliance (CSB). Research of
A.B.~Tsybakov was partially supported by the grant ANR-06-BLAN-0194
and by the PASCAL Network of Excellence. A.~Antoniadis, A.~Samarov,
and A.B.~Tsybakov  would like to thank U.~Amato for his hospitality
while visiting the  Istituto per le Applicazioni del Calcolo
`M.~Picone' CNR, Napoli to carry out this work. U.~Amato would like
to thank A.~Antoniadis for his excellent hospitality while visiting
the Laboratoire J.~Kuntzmann, Grenoble. }
}
\date{June 9, 2009}

\maketitle


\begin{abstract}
We consider the problem of multivariate density estimation when the
unknown density is assumed to follow a particular form of
dimensionality reduction, a noisy independent factor analysis (IFA)
model. In this model the data are generated by a number of latent
independent components having unknown distributions and are observed
in Gaussian noise. We do not assume that either the number of
components or the matrix mixing the components are known.
 We show that the densities of this form can be estimated with a fast
rate. Using the mirror averaging aggregation algorithm, we construct
a density estimator which achieves a nearly parametric rate
$(\log^{1/4}{n})/\sqrt{n}$, independent of the dimensionality of the
data, as the sample size $n$ tends to infinity. This estimator is
adaptive to the number of components, their distributions and the
mixing matrix. We then apply this density estimator to construct
nonparametric plug-in classifiers and show that they achieve the
best obtainable rate of the excess Bayes risk, to within a
logarithmic factor independent of the dimension of the data.
Applications of this classifier to simulated data sets and to real
data from a remote sensing experiment show promising results.




\end{abstract}



\section{Introduction}

Complex data sets lying in multidimensional spaces are a commonplace
occurrence in many areas of science and engineering. There are
various sources of this kind of data, including biology (genetic
networks, gene expression microarrays, molecular imaging data),
communications (internet data, cell phone networks), risk
management, and many others.  One of the important challenges of the
analysis of such data is to reduce its dimensionality in order to
identify and visualize its structure.

It is well known that
common nonparametric density estimators are quite
unreliable even for moderately high-dimensional data. This motivates
the use of dimensionality reduction models.
%
%
The literature on dimensionality reduction is very extensive, and we
mention here only some recent publications that are connected to our
context and contain further references (Roweis and Saul
2000;
Tenebaum, de Silva and Langford 2000;
Cook and Li 2002,
Blanchard et al.\ 2006;
Samarov and Tsybakov 2007).

In this paper we consider the independent factor analysis (IFA)
model, which generalizes the ordinary factor analysis  (FA),
principal component analysis (PCA), and independent component
analysis (ICA). The IFA model was introduced by Attias
(1999)
as
a method for recovering independent hidden sources from their
observed mixtures. 
In the ordinary FA and PCA, the hidden sources are assumed to be
uncorrelated and the analysis is based on the covariance matrices,
while IFA assumes that the hidden sources (factors) are independent
and have unknown, non-Gaussian distributions.  The ICA, in its
standard form, assumes that the number of sources is equal to the
number of observed variables and that the mixtures are observed
without noise. Mixing of sources in realistic situations, however,
generally involves noise and different numbers of sources (factors)
and observed variables, and the IFA model allows for both of these
extensions of ICA.

Most of the existing ICA algorithms concentrate on recovering the
mixing matrix and either assume the known distribution of sources or
allow for their limited, parametric flexibility, see
Hyvarinen, Karhunen and Oja (2001).
Attias
(1999)
and more recent IFA
papers
(An, Xu and Xu 2006;
Montanari, Cal\`o and Viroli 2008) either use mixture of Gaussian
distributions as source models or assume that the number of
independent sources is known, or both. In the present paper the IFA
serves as a dimensionality reduction model for multivariate
nonparametric density estimation; we suppose that the distribution
of the sources (factors)
and their number are unknown. 

Samarov and Tsybakov
(2004) 
 have shown
that densities which have the standard, noiseless ICA representation
can be estimated at an optimal one-dimensional nonparametric rate,
without knowing the mixing matrix of the independent sources.  Here
our goal is to estimate a multivariate density in the noisy IFA
model with unknown number of latent independent components observed
in Gaussian noise.  It turns out that the density generated by this
model can be estimated with a very fast rate. In Section~\ref{sec
model} we show that, using recently developed methods of aggregation
(Juditsky et al.\ 2005, 2008), we can estimate the density of this
form at a parametric root-$n$ rate, up to a logarithmic factor
independent of the dimension $d$.

One of the main applications of multivariate density estimators is
in the supervised learning. They can be used to construct plug-in
classifiers by estimating the densities of each labeled class.
Recently, Audibert and Tsybakov
(2007) have shown that plug-in classifiers can achieve fast rates of
the excess Bayes risk and under certain conditions perform better
than classifiers based on the (penalized) empirical risk
minimization. A difficulty with such density-based plug-in
classifiers is that, even when the dimension $d$ is moderately
large, most density estimators have poor accuracy in the tails,
i.e., in the region which is important for classification purposes.
Amato, Antoniadis and Gr\'egoire (2003) have suggested to overcome
this problem using the ICA model for multivariate data. 
The resulting method appears to outperform linear, quadratic and
flexible discriminant analysis
(Hastie, Tibshirani and Buja 1994) in the training set, but its
performance is rather poor in the testing set. Earlier, Polzehl
(1995) suggested a discrimination-oriented version of projection
pursuit density estimation, which appears to produce quite good
results but at a high computational cost. His procedure depends on
some tuning steps, such as bandwidth selection, which are left open
and appear to be crucial for the implementation.
More recently, Montanari et al.\ (2008) constructed plug-in
classifiers based on the IFA model, with the sources assumed to be
distributed according to a mixture of Gaussian distributions,  and
reported promising numerical results.

In Section \ref{sec:classi} we give a bound to the excess risk of
nonparametric plug-in classifiers in terms of the MISE of the
density estimators of each class. Combining this bound with the
results of Section \ref{sec model}, we show that if the data in each
class are generated by a noisy IFA model, the corresponding plug-in
classifiers achieve, within a logarithmic factor independent of the
dimensionality $d$, the best obtainable rate of the excess Bayes
risk. In Section \ref{sec:algor} we describe the algorithm
implementing our classifier. Section \ref{sec:examples} reports
results of the application of the algorithm to simulated and real
data.

\section{Independent factor analysis model for density estimation} \label{sec model}

We consider the noisy IFA model:
\begin{equation}\label{noisyICA}
\mathbf{X}= A\mathbf{S} + \bepsilon,
\end{equation}
where $A$ is a $d \times m$ unknown deterministic matrix of factor
loadings with unknown $m < d$, $\mathbf{S}$ is an unobserved
$m$-dimensional random vector with independent zero-mean components
(called factors) having unknown distributions each admitting a
density and a finite variance, and $\bepsilon$ is a random  vector
of noise, independent of $\mathbf{S}$, which we will assume to have
$d$-dimensional normal distribution with zero mean and covariance
matrix $\sigma^2 \mathbf{I}_d$, $\sigma^2>0$. Here $\mathbf{I}_d$
denotes the $d \times d$ identity matrix.

Assume that we have independent observations $\mathbf{X}_1, \dots ,
\mathbf{X}_n$,
 where each $\mathbf{X}_i$ has the same
distribution as $\mathbf{X}$. As mentioned in the Introduction, this
model is an extension of
 the ICA model, which is widely used in signal processing for blind
 source
 separation. In the signal processing literature the components of
 ${\bf S}$ are called sources rather than factors. The basic ICA
model assumes $\bepsilon=0$  and $m=d$ (cf., e.g.,
Hyvarinen et al.\ 2001). Unlike in the signal processing literature,
our goal here is to estimate the target density
$p_{\mathbf{X}}(\cdot)$ of $\mathbf{X}$, and model (\ref{noisyICA})
serves as a particular form of dimensionality reduction for density
estimation.

 Somewhat different versions of
this model where the signal $\mathbf{S}$ has not necessarily
independent components and needs to be non-Gaussian were considered
recently by
Blanchard et al.\ (2006),  Samarov and Tsybakov
(2007).
Blanchard et al.\ (2006) and the follow-up paper by
Kawanabe et al.\ (2007) use projection pursuit type techniques to
identify the non-Gaussian subspace spanned by the columns of $A$
with known number of columns $m$, while Samarov and Tsybakov
(2007) propose aggregation methods to estimate the density of
$\mathbf{X}$ when neither the non-Gaussian subspace, nor its
dimension are known.


It is well known that the standard, covariance-based factor analysis
model is not fully identifiable without extra assumptions (see,
e.g.,
Anderson and Rubin 1956). Indeed, the factors are defined only up to
an arbitrary rotation. The independence of factors assumed in
(\ref{noisyICA}) excludes this indeterminacy provided that at
most one factor is allowed to have a Gaussian distribution. This
last assumption is standard in the ICA literature and we will also
make it throughout the paper.
We will also assume throughout that
the columns of $A$ are orthonormal.

By independence between the noise and the vector of factors ${\bf
S}$, the target density $p_{\mathbf{X}}$ can be written as a
convolution:
\begin{equation} \label{conv1}
p_{\mathbf{X}}(\mathbf{x}) = \int_{\RR^m} p_{\mathbf{S}}(\mathbf{s})
\phi_{d,\sigma^2}(\mathbf{x}-A\mathbf{s}) d\mathbf{s},
\end{equation}
where $\phi_{d,\sigma^2}$ denotes the density of a $d$-dimensional
Gaussian distribution $N_d(0,\sigma^2\mathbf{I}_d)$.

Since  in (\ref{conv1}) we have a convolution with a Gaussian
distribution, the density $p_{\mathbf{X}}$ has very strong
smoothness properties, no matter how irregular the density
$p_{\mathbf{S}}$ of the factors is, whether or not the factors are
independent, and whether or not the mixing matrix $A$ is known. In
the Appendix, we construct a kernel estimator $\hat{p}^*_n$ of
$p_{\mathbf{X}}$ such that
\begin{equation}\label{rate1}
\EE||\hat{p}^*_n-p_{\mathbf{X}}||^2_2 \le C \frac{(\log
n)^{d/2}}{n},
\end{equation}
where $C$ is a constant and $||\cdot||_2$ is the $L_2(\RR^d)$ norm.
As in  Artiles
(2001),
Belitser and Levit (2001), it is not hard to show that the rate
given in (\ref{rate1}) is optimal for the class of densities
$p_{\mathbf{X}}$ defined by (\ref{conv1}) with arbitrary
$p_{\mathbf{S}}$.

Though this rate appears to be very fast asymptotically, it does not
guarantee good accuracy for most practical values of $n$, even if
$d$ is moderately large. For example, if $d=10$, we have $(\log
n)^{d/2}>n$ for all $n\le 10^5$. As we show below, the assumed
independence of the sources and orthogonality of $A$ allows us to
eliminate the dependence of the rate on the dimension $d$.

In order to construct our estimator, we first consider the
estimation of $p_{\mathbf{X}}$ when the dimension $m$, the mixing
matrix $A$, and the level of noise $\sigma^2$ are specified; the
fact that none of these quantities is known is addressed later in
this section.

Since the columns of $A$ are orthonormal, we have
$A^T\mathbf{X}={\bf S}+A^T\bepsilon$ and
\begin{eqnarray*}
 \phi_{d,\sigma^2}(\mathbf{x}-A\mathbf{s}) & = &  \left( \frac{1}{2\pi \sigma^2}\right)^{d/2}
 \exp \left\{ - \frac{1}{2\sigma^2}
 (\mathbf{x}-A\mathbf{s})^T (\mathbf{x}-A\mathbf{s}) \right\} \\
 & = & \left( \frac{1}{2\pi \sigma^2}\right)^{d/2} \exp \left\{ - \frac{1}{2\sigma^2}
 (\mathbf{s}-A^T\mathbf{x})^T(\mathbf{s}-A^T\mathbf{x}) \right\} \cdot \exp\left\{
- \frac{1}{2\sigma^2} \mathbf{x}^T (\mathbf{I}_d - A A^T) \mathbf{x}
\right\}.
\end{eqnarray*}
Substitution of the above expression in  (\ref{conv1}) gives:
\begin{equation*}
p_{\mathbf{X}}(\mathbf{x}) =
 \left( \frac{1}{2\pi \sigma^2}\right)^{(d-m)/2}  \exp\left\{
- \frac{1}{2\sigma^2} \mathbf{x}^T (\mathbf{I}_d - A A^T) \mathbf{x}
\right\} \int_{\RR^m} p_{\mathbf{S}}(\mathbf{s})
\phi_{m,\sigma^2}(\mathbf{s}-A^T\mathbf{x}) d\mathbf{s}.
\end{equation*}
Now, by independence of the factors, we get:
\begin{equation}\label{conv3}
p_{\mathbf{X}}(\mathbf{x}) \equiv p_{m,A}(\mathbf{x}) =
 \left( \frac{1}{2\pi \sigma^2}\right)^{(d-m)/2}  \exp\left\{
- \frac{1}{2\sigma^2} \mathbf{x}^T (\mathbf{I}_d - A A^T) \mathbf{x}
\right\} \prod_{k=1}^m g_k(\mathbf{a}_k^T \mathbf{x})
\end{equation}
 where $\mathbf{a}_k$ denotes the $k$th column of $A$ and
\begin{equation}\label{gk}
g_k(u) = (p_{S_k} \ast \phi_{1,\sigma^2})(u) = \int_{\RR} p_{S_k}
(s) \phi_{1,\sigma^2}(u-s) ds.
\end{equation}
We see that to estimate the target density $p_{\mathbf{X}}$ it
suffices to estimate nonparametrically each one-dimensional density
$g_k$ using the projections of an observed sample $\mathbf{X}_1,
\dots , \mathbf{X}_n$ generated by the model~(\ref{noisyICA})  onto
the $k$th direction $\mathbf{a}_k$.

Note that, similarly to (\ref{conv1}), the density $g_k$ is obtained
from convolution with a one-dimensional Gaussian density, and
therefore has very strong smoothness properties. To estimate $g_k$
we will use the kernel estimators
\begin{equation} \label{kern1}
 \hat{g}_{k}(x) = \frac{1}{nh_n} \sum_{i=1}^n K\left( \frac{x -
 \mathbf{a}_k^T
 \mathbf{X}_i}{h_n}\right), \quad k=1,...,m,
 \end{equation}
with a bandwidth $h_n \asymp (\log n)^{-1/2}$ and the sinc function
kernel $K(u) = \sin u / \pi u $. We could also use here any other
kernel $K$ whose Fourier transform is bounded and compactly
supported, for example, the de la Vall\'ee-Poussin kernel  $K (u) =
(\cos(u)- \cos(2u))/(\pi u^2)$, which is absolutely integrable and
therefore well suited for studying the $L_1$-error.

A potential problem of negative values of $\hat{g}_{k}$ in the
regions where the data are sparse can be corrected using several
methods (see, for example,
Hall and Murison 1993; Glad, Hjort and Ushakov 2003).
For our practical implementation
we will follow the method suggested in
Hall and Murison (1993),
and our estimators
will be obtained by truncating the estimator $\hat{g}_{k}(x)$
outside the ``central'' range where it is nonnegative,  and then
renormalizing.

Once each ``projection'' density $g_k$ is estimated by the
corresponding kernel estimator  ~(\ref{kern1}), the full target
density $p_{\mathbf{X}}$ is then estimated using (\ref{conv3}):
\begin{equation}\label{convest}
\hat{p}_{n,m,A}(\mathbf{x}) =
 \left( \frac{1}{2\pi \sigma^2}\right)^{(d-m)/2}  \exp\left\{
- \frac{1}{2\sigma^2} \mathbf{x}^T (\mathbf{I}_d - A A^T) \mathbf{x}
\right\} \prod_{k=1}^m \hat{g}_{k}(\mathbf{a}_k^T \mathbf{x}).
\end{equation}

The following proposition proved in the Appendix summarizes the
discussion for the case when $A$ and $\sigma^2$ are known.

\begin{prop}\label{prop1} Consider a random sample of size $n$ from the density $p_{\mathbf{X}}$ given
by (\ref{conv3}) with known $A$ and $\sigma^2$. Then the estimator
(\ref{convest}) with $\hat{g}_{k}$ given in (\ref{kern1}) has the
mean integrated square error of the order $(\log n)^{1/2}/n$:
\begin{equation}
\label{mise} \EE \| \hat{p}_{n,m,A} - p_{\mathbf{X}}\|_2^2 =
\mathcal{O}\left( \frac{(\log n)^{1/2}}{n}\right).
\end{equation}

\end{prop}
Note that neither $m$ nor $d$ affect the rate. Note also that
Proposition \ref{prop1} is valid with no assumption on the
distribution of the factors. The identifiability assumption (that at
most one factor is allowed to have a Gaussian distribution) is not
used in the proof, since we do not estimate the matrix $A$.

\medskip

So far in this section we have assumed that $A$ and $\sigma^2$ are
known. When $\sigma^2$ is an unknown parameter, it is still possible
to obtain the same rates based on the approach outlined above,
provided that the dimensionality reduction holds in the strict
sense, i.e., $m<d$. Indeed, assume that we know an upper bound $M$
for the number of factors $m$ and that $M < d$. For example, if the
dimensionality reduction in the strict sense holds, we can take
$M=d-1$. The assumption $M < d$ is only needed to estimate the
variance of the noise; if $\sigma^2$ is known we allow $M=d$.

The assumed independence and finite variance of the factors imply
that their covariance matrix, which we will denote by $W$, is
diagonal. The covariance matrix $\Sigma_{\mathbf{X}}$ of
$\mathbf{X}$ is given by:
$$ \Sigma_{\mathbf{X}} = A W A^T + \sigma^2 \mathbf{I}_d.$$
If $\lambda_1(\Sigma_{\mathbf{X}})\ge \cdots \ge
\lambda_d(\Sigma_{\mathbf{X}})$ denote the eigenvalues of
$\Sigma_{\mathbf{X}}$ sorted in decreasing order, then
$\lambda_i(\Sigma_{\mathbf{X}})=w_i + \sigma^2$, for $i=1,\dots, m$,
and $\lambda_i(\Sigma_{\mathbf{X}})=\sigma^2$ for $i>m$, where $w_i$
 denote the diagonal elements of $W$.
 We estimate $\sigma^2$ with
$$\hat{\sigma}^2 = \frac{1}{d-M} \sum_{i=M+1}^d \hat{\lambda}_i,$$
where $\hat{\lambda}_i$, $i=1, \dots, d$, are the eigenvalues of the
sample covariance matrix $\hat{\Sigma}_{\mathbf{X}}$ arranged in
decreasing order. Note that $\hat{\sigma}^2$ is a root-$n$
consistent estimator. Indeed, the root-$n$ consistency of each
$\hat{\lambda}_i$ is a consequence of elementwise root-$n$
consistency of $\hat{\Sigma}_{\mathbf{X}}$ and of the inequality
$$|\lambda_i(C+D)-\lambda_i(C)| \le \|D\|_2, \quad i=1,2,...,d,$$
where $C$ and $D$ are any symmetric matrices and $\|D\|_2$ is the
spectral norm of~$D$. The last inequality easily follows from a
classical inequality of Fan (1951).

 Using the root-$n$ consistency of $\hat{\sigma}^2$, it is not hard to show that the estimation of
$\sigma^2$ does not affect a slower density estimator rate, and so in
what follows we will assume that $\sigma^2$ is known.

\medskip

Consider now the case where the index matrix $A$, and hence its rank
$m$, are unknown. We will use a model selection type aggregation
procedure similar to the one developed recently by Samarov and
Tsybakov
(2007) and, more specifically, the mirror averaging algorithm of
Juditsky, Rigollet and Tsybakov
(2008). We aggregate estimators of the type (\ref{convest})
corresponding to candidate pairs $(k,{\hat B}_k)$, $k = 1,\dots, M$.
Here ${\hat B}_k$ is
a $d \times k$ matrix whose columns are the first $k$ (in the
decreasing order of eigenvalues) orthonormal eigenvectors of the
spectral decomposition of $\hat{\Sigma}_{\mathbf{X}} -
\hat{\sigma}^2 {\bf I}_d$ (and thus of
$\hat{\Sigma}_{\mathbf{X}}$).
For the true rank $m$, it follows from Lemma A.1 of Kneip and Utikal (2001) that, provided
that $m$ largest eigenvalues of $\Sigma_{\mathbf{X}}-\sigma^2{\bf I}_d$ are distinct and positive
and the 4th moments of the components of ${\bf X}$ are finite,
${\hat B}_m$ is a $\sqrt{n}$-consistent estimator of $A$.

We can now define the aggregate estimator, applying the results of
Juditsky, Rigollet and Tsybakov
(2008) in our framework. We split the sample $\mathbf{X}_1$, \dots,
$\mathbf{X}_n$ in two parts, ${\cal D}_1$ and ${\cal D}_2$ with $n_1
= \hbox{Card}({\cal D}_1)$, $n_2 = \hbox{Card}({\cal D}_2)$,
$n=n_1+n_2$. From the first subsample ${\cal D}_1$ we construct the
estimators
\begin{equation}\label{estk}
\hat{p}_k(\mathbf{x}) \equiv \hat{p}_{n_1,k,{\hat B}_k}(\mathbf{x})
=
 \left( \frac{1}{2\pi \sigma^2}\right)^{(d-k)/2}  \exp\left\{
- \frac{1}{2\sigma^2} \mathbf{x}^T (\mathbf{I}_d - {\hat B}_k {\hat
B}_k^T) \mathbf{x} \right\} \prod_{j=1}^k \hat{g}_{j}(
\mathbf{b}_{k,j}^T \mathbf{x})
\end{equation}
for $k=1,\dots, M$, where $\mathbf{b}_{k,j}$ denotes the $j$th
column of ${\hat B}_k$, the estimators $\hat{g}_{j}(\cdot)$ are
defined in (\ref{kern1}), and both ${\hat B}_k$ and
$\hat{g}_{j}(\cdot)$ are based only on the first subsample ${\cal
D}_1$.

The collection $\mathcal{C}$ of density estimators $\left\{
\hat{p}_{n_1,k,{\hat B}_k},\ k = 1,\dots,M \right\}$ of the form
(\ref{estk})
 constructed from the subsample ${\cal D}_1$ can be considered as a collection of fixed
functions when referring to the second subsample ${\cal D}_2$. The
cardinality of this collection is $M$.

To proceed further, we need some more notation. Let $\Theta$ be the
simplex
$$\Theta = \left\{ \boldsymbol{\theta} \in \RR^M\, : \sum_{k=1}^M \theta_k =1, \  \theta_k \geq 0, \  k=1, \dots, M \right\}, $$
and
$$\mathbf{u}(\mathbf{X}) = \left(u_1(\mathbf{X}), \dots, u_M(\mathbf{X})\right)^T,$$
where
\begin{equation}
u_k(\mathbf{x}) = \int \hat{p}_k^2(\mathbf{x}) d\mathbf{x} - 2
\hat{p}_k(\mathbf{x}). \label{uk}
\end{equation}
Introduce the vector function
$$\mathbf{H}(\mathbf{x}) = \left(\hat{p}_1(\mathbf{x}), \dots,
\hat{p}_M(\mathbf{x})\right)^T.$$ As in
Juditsky, Rigollet and Tsybakov (2008), the goal of aggregation is
to construct a new density estimator $\tilde{p}_n(\mathbf{x})$
 of the form
\begin{equation}
\tilde p_n(\mathbf{x})=\tilde{\btheta}^T \mathbf{H} (\mathbf{x})
\label{recursive}
\end{equation}
which is nearly as good in terms of the $L_2$-risk as the best one
in the collection $\mathcal{C}$. Using the mirror averaging
algorithm, the aggregate weights $\tilde{\btheta}$ are computed by a
simple procedure which is recursive over the data. Starting with an
arbitrary value $\tilde{\btheta}^{(0)} \in \Theta$, these weights
are defined in the form:
\begin{equation}
\tilde{\btheta} = \frac{1}{n_2}\sum_{\ell=1}^{n_2} \tilde{\btheta}^{(\ell-1)},
\label{theta1}
\end{equation}
where the components of
$\tilde{\btheta}^{(\ell)}$
are given by
\begin{equation}
\tilde{\theta}_k^{(\ell)} = \frac
{
\exp \left( -\beta^{-1} \sum_{r=1}^\ell u_k(\mathbf{X}_r) \right)
}
{
\sum_{t=1}^M \exp \left( -\beta^{-1} \sum_{r=1}^\ell u_t(\mathbf{X}_r) \right)
}
,\ k=1,\ldots, M,
\label{theta2}
\end{equation}
with $\mathbf{X}_r$, $r=1,\dots, n_2$, denoting the elements of the
second subsample ${\cal D}_2$. Here $\beta>0$ is a random variable
measurable w.r.t. the first subsample ${\cal D}_1$.

Our main result about the convergence of the aggregated density
estimator is given in Theorem \ref{th1} below. We will consider the
norms restricted to a Euclidean ball $B \subset \RR^d$:
$\|f\|_{2,B}^2=\int_{B} f^2(\mathbf{x})d\mathbf{x}$,
$\|f\|_{\infty,B}=\sup_{t\in B} |f(t)|$ for $f:\RR^d\to \RR$.
Accordingly, in Theorem \ref{th1} we will restrict our estimators to
$B$ and define $\tilde{p}_n$ by the above aggregation procedure
where $\hat{p}_k(x)$ are replaced by $\hat{p}_k(x)I\{x\in B\}$. Here
$I\{\cdot\}$ denotes the indicator function.

Clearly, all densities ${p}_{\mathbf{X}}$ of the form (\ref{conv3})
are bounded: $\|p_{\mathbf{X}}\|_{\infty,B}\le L_0:=(2\pi
\sigma^2)^{-d/2}$ for all $m$ and $A$.
We set ${\hat L}_1=\max_{k=1,\dots,M}\|\hat{p}_k\|_{\infty,B}$ and
${\hat L}=\max(L_0,{\hat L}_1)$. In the Appendix we prove that
\begin{equation}\label{bdd1}
\EE \|\hat{p}_k\|_{\infty,B} \le L', \quad \forall k=1,...,M,
\end{equation}
where $L'$ is a constant.

\begin{theo} \label{th1} Let $p_\mathbf{X}$ be  the density of ${\bf X}$
in model~(\ref{noisyICA}).  Assume that covariance matrix
$\Sigma_{\mathbf{X}}$ has distinct eigenvalues and the 4th moments
of the components of ${\bf X}$ are finite. Let $n_2 = [
{cn}/{\sqrt{\log n}}]$ for some constant $c>0$ such that $1 \leq n_2
< n$. Then for $\beta=12{\hat L}$, the
aggregate estimator $\tilde{p}_n$ with $\tilde{\btheta}$ obtained by
the mirror averaging algorithm restricted to a Euclidean ball $B$
satisfies
\begin{equation}\label{th1e}
\EE \| \tilde{p}_n - p_\mathbf{X}\|_{2,B}^2 =
\mathcal{O}\left(\frac{(\log n)^{1/2}}{n}\right), \end{equation} as
$n \rightarrow +\infty$.
\end{theo}

\smallskip

 The theorem implies that the estimator
$\tilde{p}_n$ adapts to the unknown $m$ and $A$, i.e., has the same
rate, independent of $m$ and $d$, as in the case when the dimension
$m$ and the matrix $A$ are known. The proof is given in the
Appendix.

{\bf Remarks}.

1. Inspection of the proof shows that Theorem \ref{th1} holds with
no assumption on distributions of the factors (except that at
most one of them can be Gaussian). In particular, we do not need
them to have densities with respect to the Lebesgue measure.

 2. We state Theorem \ref{th1} with a restricted $L_2$-norm
$\| \cdot \|_{2,B}$. Under mild assumptions on the densities of the factors we can
extend it to the $L_2$-norm on $\RR^d$. Indeed, inspection of the
proof shows that Theorem \ref{th1} remains valid for balls $B$ of
radius $r_n$ which tends to infinity slowly enough as $n\to \infty$.
If $p_{\bf X}$ behaves itself far from the origin roughly as a
Gaussian density (which is true under mild assumptions on factor densities), then the integral of $p_{\bf X}^2$ outside of the ball reduces
to a value smaller than the right hand side of (\ref{th1e}).

\section{Application to nonparametric classification}\label{sec:classi}

One of the main applications of multivariate density estimators is
in the supervised learning, where they can be used to construct
plug-in classifiers by estimating the densities of each labeled
class. The difficulty with such density-based plug-in classifiers is
that, even for moderately large dimensions $d$, standard density
estimators have poor accuracy in the tails, i.e., in the region
which is important for classification purposes. In this section we
consider the nonparametric classification problem and bound the
excess misclassification error of a plug-in classifier in terms of
the MISE of class-conditional density estimators. This bound implies
that, for the class-conditional densities obeying the noisy IFA
model~(\ref{conv1}), the resulting plug-in classifier has nearly
optimal excess error.

Assume that we have $J$ independent training samples
$\{X_{j1},\dots,X_{jN_j}\}$ of sizes $N_j$, $j=1,\dots, J$, from $J$
populations with densities $f_1,\dots,f_J$ on $\RR^d$. We will
denote by $\cal D$ the union of training samples. Assume that we
also have an observation $\mathbf{X}\in\RR^d$ independent of these
samples and distributed according to one of the $f_j$. The
classification problem consists in predicting the corresponding
value of the class label $j\in \{1,\dots,J\}$. We define a
classifier or prediction rule   as a measurable function $T(\cdot)$
which assigns a class membership based on the explanatory variable,
i.e., $T : \RR^d \rightarrow \{1,\dots,J\}.$ The misclassification
error associated with a classifier $T$ is usually defined as
$$R(T)= \sum_{j=1}^J \pi_j \PP_j(T(\mathbf{X}) \not= j)=
\sum_{j=1}^J \pi_j \int_{\RR^d} I( T(\mathbf{x}) \not=j)
f_j(\mathbf{x}) d \mathbf{x}
$$
where $\PP_j$ denotes the class-conditional population probability
distribution with density $f_j$, and $\pi_j$ is the prior
probability of class $j$. We will consider a slightly more general
definition:
$$R_B(T)=
\sum_{j=1}^J \pi_j \int_{B} I( T(\mathbf{x}) \not=j) f_j(\mathbf{x})
d \mathbf{x}
$$
where $B$ is a Borel subset of $\RR^d$. The Bayes classifier $T^*$
is the one with the smallest misclassification error:
$$R_B(T^{*}) = \min_T R_B(T).$$
In general, the Bayes classifier is not unique. It is easy to see
that there exists a Bayes classifier  $T^*$ which does not depend on
$B$ and which is defined by
$$
\pi_{T^*(\mathbf{x})} f_{T^*(\mathbf{x})}(\mathbf{x})= \min_{1\le j
\le J}\pi_j f_j(\mathbf{x}), \quad \forall \ \mathbf{x}\in \RR^d.
$$
A classifier trained on the sample ${\cal D}$  will be denoted by
$T_{\cal D}({\bf x})$. A key characteristic of such a classifier is
the misclassification  error $R_B(T_{\cal D})$. One of the main
goals in statistical learning is to construct a classifier with the
smallest possible excess risk
$$\mathcal{E}(T_{\cal D}) = \EE R_B(T_{\cal D}) - R_B(T^{*}).$$
We consider plug-in classifiers $\hat{T}(\mathbf{x})=\hat{T}_{\cal
D}(\mathbf{x})$ defined by:
$$
\pi_{\hat{T}(\mathbf{x})} \hat{f}_{\hat{T}(\mathbf{x})}(\mathbf{x})=
\min_{1\le j \le J}\pi_j \hat{f}_{j}(\mathbf{x}), \quad \forall \
\mathbf{x}\in \RR^d
$$
where $ \hat{f}_{j}$ is an estimator of density $f_j$ based on the
training sample $\{X_{j1},\dots,X_{jN_j}\}$.

The following proposition relates the excess risk
$\mathcal{E}(\hat{T})$ of plug-in classifiers to the rate of
convergence of the estimators $\hat{f}_{j}$.
\begin{prop} \label{prop2}
$$\mathcal{E}(\hat{T})  \leq \sum_{j=1}^J \pi_j\,
\EE \int_B |\hat{f}_{j}(\mathbf{x}) - f_j(\mathbf{x}) | d\mathbf{x}
$$
\end{prop}
Proof of the proposition is given in the Appendix.

Assume now that  the class-conditional densities follow the noisy
IFA model (\ref{conv1}) with different unknown mixing matrices and
that $N_j \asymp n$ for all $j$. Let $B$ be a Euclidean ball in
$\RR^d$ and define each of the estimators $\hat{f}_{j}$ using the
mirror averaging procedure as in the previous section. Then, using
Theorem \ref{th1}, we have
$$
\EE \int_B |\hat{f}_{j}(\mathbf{x}) - f_j(\mathbf{x}) | d\mathbf{x}
\le \sqrt{|B|} \ \EE \| \hat{f}_{j} - f_j\|_{2,B} =
\mathcal{O}\left(\frac{(\log n)^{1/4}}{\sqrt{n}}\right)
$$
as $n\to\infty$, where $|B|$ denotes the volume of the ball $B$.
Thus, the excess risk $\mathcal{E}(\hat{T})$ converges to 0 at the
rate $(\log n)^{1/4}/\sqrt{n}$ independently of the dimension $d$.
Following the argument in Devroye, Gy\"orfi and Lugosi (1996) or
Yang (1999), it is easy to show that this is the best obtainable
rate for the excess risk, up to the $\log^{1/4} n$ factor.

\section{The algorithm}\label{sec:algor}
In this section we discuss numerical aspects of the proposed density
estimator.

Clearly, one-dimensional kernel density estimators $\hat{g}_k$ with
given bandwidth, say $h_n \propto (\log n)^{-1/2}$, can be computed
in a fast way. Similarly, estimating the variance of the noise
component in the noisy IFA model amounts to implementing a single
singular value decomposition (SVD) of the $d \times n$ data matrix
$D=({\bf X}_1, \dots, {\bf X}_n)$. Let $D=V\Lambda U^T$ be the SVD
of $D$, where $\Lambda$ is the diagonal matrix and $U$, $V$ are
matrices with orthonormal columns. We assume w.l.o.g. that ${\bf
X}_i$ are centered. Then an estimate of the variance
$\hat{\sigma}^2_k$ with rank $k$ approximation, $k\le M$, is given
by
\begin{equation}
\hat\sigma^2_k=\frac1{d-k} \sum_{i=k+1}^d s^2_i, \quad
k=1,\dots,M\label{sigmak}
\end{equation}
where $s_i$ are the diagonal elements of $\Lambda/\sqrt{n}$
sorted in the decreasing
order. When the index matrix $A$ is unknown, the rank $k$
approximation ${\hat B}_k$ of $A$ used in the density estimator
$\hat{p}_{k}$, cf. (\ref{estk}), can be easily obtained from the SVD
of $D$. Indeed, we can take ${\hat B}_k=V_k$, where $V_k$ is formed
by the first $k$ columns of $V$. So, accurate computation of the
density estimators (\ref{estk}) is feasible, reasonably fast and
does not require a huge amount of memory even for very large $n$ and
$d$.

Therefore, the complexity of the procedure is controlled by the
numerical implementation of the mirror averaging algorithm which, in
particular, requires the computation of the score functions
$u_k(\mathbf{x})$, involving integration of $\hat{p}^2_k$, see
(\ref{uk}). The numerical implementation of the integral of the
square of density estimates $\hat{p}_k$ in $\RR^d$ can be realized
by means of cubature formulas. Recall that for the calculation of
$\int \hat{p}_k({\bf x})^2 d{\bf x}$, say, a cubature has the form
$\sum_{i=1}^N w_i \hat{p}_k^2({\bf x}_i)$ where ${\bf x}_i$ are the
nodes and $w_i$ are the associated weights. In our setting, $M$
integrals involving the ${\hat B}_k$-projections need to be
calculated for each $\theta_k$, so formulas with fixed nodes will be
actually more economical. On multidimensional domains, product
quadratures quickly become prohibitive (they grow exponentially in
$d$ for the same accuracy), and therefore this approach is not
realistic.

An alternative is to use Monte-Carlo integration methods which
require much more evaluations but do not depend on the dimension
$d$, or a more clever implementation through Gibbs sampling  by
generating samples from some suitable distribution for the
Monte-Carlo estimates. Several Gibbs sampling strategies were
considered in the present work. The fastest one was to generate
samples directly from $\hat{p}_k$, so that
\begin{equation*}
\int \hat{p}^2_k({\bf x})d{\bf x} \simeq \frac 1Q \sum_{i=1}^Q
\hat{p}_k({\bf x}_i),
\end{equation*}
where $Q$ is the number of generated i.i.d.\ random realizations
${\bf x}_i$ from the density $\hat{p}_k$.

The overall algorithm implementing our approach is the following:

\begin{algorithm}
{\rm
\begin{description}
\item{-} Compute the singular value decomposition of the data array
$D$:
\begin{equation*}
D=V\Lambda U^T,
\end{equation*}
with matrices $U$, $V$, and $\Lambda$ having dimensions $n \times
d$, $d \times d$ and $d \times d$, respectively;

\item{-} {\tt for} $k${\tt=}$1${\tt ,}$\ldots${\tt ,}$M$

\begin{description}
\item{} Take ${\hat B}_k$ as the matrix built from the first $k$ columns of $V$;

\item{} Compute $\hat\sigma^2_k$ from  (\ref{sigmak});

\item{} Compute the density estimator $\hat{p}_k(\mathbf{x})$ from
(\ref{estk}) based on the subsample ${\cal D}_1$ ;

\item{} Compute $u_k(\mathbf{x})$ from (\ref{uk}).
\end{description}

\item{-} {\tt end for}
\item{-}
Estimate the weights through  (\ref{theta1})--(\ref{theta2}) and
output the final density estimator  (\ref{recursive}).
\end{description}
}
\end{algorithm}
To speed up computations, one-dimensional kernel density estimators
$\hat{g}_j$, $j=1,\ldots,M$, in (\ref{estk}) are obtained through a
Fast Fourier Transform algorithm, cf.
Silverman (1982).

The algorithm for estimating $\int \hat{p}^2_k({\bf x})d{\bf x}$ in
(10) goes through the following steps.

\begin{algorithm}
{\rm
\begin{description}
\item{-}  Generate $Q$ independent random numbers, $y_k^{(i)}$, $i=1,\ldots,Q$, from each ${\hat g}_k$, $k=1,\ldots,M$, and compute the corresponding density $\hat{g}_k(y_k^{(i)})$ by kernel density estimation;
\item{-}  Generate the corresponding $d$-dimensional $\mathbf{x}^{(i)}$ as $\mathbf{x}^{(i)}={\hat B}_k \mathbf{y}^{(i)}+(I_d-\hat B_k \hat B_k^T)\bepsilon^{(i)}$,
 $\mathbf{y}^{(i)}\equiv(y_1^{(i)},\ldots,y_k^{(i)})$,  with $\bepsilon^{(i)}$ being random numbers extracted from a $d$-variate Gaussian density function having 0 mean and diagonal covariance $\hat\sigma_k^2 I_d$;
\item{-} Compute $\hat{p}_k(\mathbf{x}^{(i)})$ through (\ref{estk});
\item{-} Output the estimate $\frac1Q \sum_{i=1}^Q
\hat{p}_k({\bf x}^{(i)})$ of the integral $\int \hat{p}^2_k({\bf
x})d{\bf x}$.
\end{description}
\noindent Here $Q$ is chosen so that generating more random numbers
does not change the estimated value of the integral within a
predefined tolerance. Random numbers generated from the density
estimator $\hat{g}_k$ are based on the corresponding
 cumulative functions and pre-computed on a high resolution grid with
linear interpolation. }
\end{algorithm}

\section{Simulations and examples}\label{sec:examples}

\subsection{Density estimation}

To study the performance of density estimates based on our noisy IFA
model we have conducted an extensive set of simulations. We used
data generated from a variety of source distributions, including
subgaussian and supergaussian distributions, as well as
distributions that are nearly Gaussian. We studied unimodal,
multimodal, symmetric, and nonsymmetric distributions. Table
\ref{tab test} lists the basic (one-dimensional) test densities from
which multidimensional density functions are built.


Experiments were run up to dimension $d=6$ with a number of
independent factors equal to 1 and 2. Random i.i.d.\ noise was
generated and added to the simulated signals  so that the Signal to
Noise Ratio (SNR) was equal to 3, 5 or 7. The kernels $K$ for
density estimators ${\hat g}_j$ in (\ref{estk}) were the Gaussian,
the sinc and de la Vall\'ee-Poussin kernels; the bandwidth $h$ was
chosen as $h=\sigma/\log^{1/2} n$. To obtain legitimate (i.e.,
nonnegative) density functions they
 were post-processed by the procedure of
 Hall and Murison (1993).
 The size of the sample was chosen as
 $n$=200, 300, 500, 700, 1000, 2000 and 4000. The following
criterion was used for evaluating the performance of density
estimators:
\begin{equation}
I_1 := 100\left(1-\frac {\int \left(
{p}_{\mathrm{estimated}}(\mathbf{x})- p_{\mathbf{X}}(\mathbf{x})
\right)^2 d\mathbf{x}}
{\int p^2_{\mathbf{X}}(\mathbf{x}) d\mathbf{x}}\right).
\label{index}
\end{equation}

The performance of IFA density estimation was compared with kernel
smoothing (KS)
(see, e.g., Wand and Jones, 1995) as implemented in the {\tt KS}
package available in {\tt R}. IFA density estimation has been
implemented in the {\tt MATLAB} environment and the scripts are
available upon request.
We note that KS can be effectively computed only up to $d=6$ if
the FFT algorithm is used. In contrast with this, our method has no
practical restrictions on the dimension. This is due to the use of a
proper Gibbs sampling for estimating integrals (\ref{uk}); in
addition the density estimate can be computed on any set in $\RR^d$,
not necessarily on a lattice imposed by the FFT.

We conducted numerical experiments  by generating random samples of
size $n$ from the independent components of Table \ref{tab test},
random mixing matrices, and different realizations of Gaussian
noise. In particular, the elements of the mixing matrix $A$ were
generated as i.i.d. standard Gaussian random variables and then the
matrix was orthonormalized by a Gram-Schmidt procedure. We perform
50 Monte-Carlo replications for each case and output the
corresponding values $I_1$. Results over all experiments show a very
good performance of Noisy IFA. For brevity we only show some
representative figures in the form of boxplots. We display different
test functions to demonstrate good performances over all of them.
Moreover, we present only the case of SNR=3 because it seems to be
more interesting for applications and because improvement of
performance for both methods flattens the differences. Figure
\ref{figboxplotd2f2snr3} shows the case of $d=2$, SNR=3 and test
function 2 (chi-square function), where the superiority of the
aggregated Noisy IFA with respect to KS is clear.
Figure \ref{fig boxplotd3f3snr3} shows analogous boxplots in the
case $d=3$ and test function 3 (mixture of Gaussians), again when
SNR=3. This case is interesting because the dimension $d$ is larger,
whereas the number of independent factors is kept constant with
respect to the previous experiment. Figure \ref{fig boxplotd3f3snr3}
clearly shows that difference of performance between Noisy IFA and
KS increases in favor of the former.
Finally, Figure \ref{fig boxplotd5f5_6snr3} shows boxplots in the
case $d=5$ and test functions 5 and 6 (chi-square and Student,
respectively), again for SNR=3. Better performance of Noisy IFA with
respect to KS is confirmed, especially when $d$ increases.


Finally, Table \ref{tab 5} shows typical computational times
 of aggregated IFA and KS density
estimators. Executions were run on
a single core 64-bit Opteron 248
processor with {\tt MATLAB} version R2008a, {\tt R} 2.9.0 and Linux Operating System. We see that the aggregated IFA is more than one order of magnitude
faster than KS.


\subsection{Classification: a real data example}\label{sec realdata}

In this subsection we apply the nonparametric classification method
suggested in Section~\ref{sec:classi}
 to real data. We consider only
a two-class problem and we assume that the class-conditional
distributions follow the noisy IFA model. To evaluate the
performance of our approach in comparison with other classification
methods that are often used in this context, we have also applied to
these data three other classification procedures, one parametric and
two nonparametric, namely:
\begin{description}
\item{LDA} (Linear Discriminant Analysis). Class-conditional
density functions are supposed to be Gaussian with a common
covariance matrix among classes, and the two classes are separated
by a hyperplane in $d$-dimensional space.
\item{NPDA} (Nonparametric Discriminant Analysis, Amato et al.\ 2003).
In this procedure class-conditional density functions are estimated
nonparametrically by the kernel method, assuming that the density
obeys an ICA model. The kernel functions
mentioned above in this section were considered in the experiments.
The smoothing procedure uses an asymptotic estimate of the bandwidth
and a correction for getting non-negative density estimators.%
\item{FDA} (Flexible Discriminant Analysis;
Hastie, Tibshirani and Buja 1994).
 This method is also nonparametric, but classification
is performed through an equivalent regression problem where the
regression function is estimated by the spline method.
 \end{description}

We  have compared the performance of the classification methods on a
data set from a remote sensing experiment. MSG (METEOSAT Second
Generation) is a series of geostationary satellites launched by
EUMETSAT (EUropean organization for the exploitation of
METeorological SATellites) mainly aimed at providing data useful for
the weather forecast. A primary instrument onboard MSG is SEVIRI, a
radiometer measuring radiance emitted by Earth at $d=11$ spectral
channels having a resolution of 3 Km$^2$ at sub-satellite point.
Essentially, SEVIRI produces 11 images of the whole Earth hemisphere
centered at $0^{\mathrm{o}}$ degrees latitude every 15 minutes.
Recognizing whether each pixel of the images is clear or affected by
clouds (cloud detection) is a mandatory preliminary task for any
processing of satellite data. In this respect multispectral radiance
data are prone to improve the detectability of clouds, thanks to the
peculiar behavior of clouds in selected spectral bands. Figure
\ref{fig msg} shows an RGB image of the Earth taken by SEVIRI on
June 30th 2006 UTC time 11:12 composed by 3 selected spectral
channels.
The problem of cloud detection is to infer the possible presence of
clouds for each pixel of the images. In order to accomplish this
task by discriminant analysis a training set has to be defined. Here
we take the training set from a cloud mask produced by sensor MODIS
onboard NOAA EOS series satellites. MODIS sensor is endowed with a
product (MOD35) aimed to produce a reliable cloud mask in many
pixels (confident classification in the terminology of MOD35). The
algorithm underlying MOD35 is based on physical arguments, with a
series of simple threshold tests mostly based on couples of spectral
bands (see
Platnick et al.\ (2003)
for details of the
algorithm). Troubles in dealing with the increasing number of spectral bands of current and next generation
instrumentation from the physical point of view is fostering investigation of statistical methods for detecting clouds.
Due to the very different spectral characteristics of water and land pixels, two separate independent classifications
are performed for the two cases. Over land the MOD35 data set is composed of 11289 cloudy pixels and 19022 clear ones;
for water pixels we have 14585 cloudy pixels and 16619 clear ones. We assume that labels assigned by MOD35 are the truth.

 In order to evaluate the methods, for each case (land and water) we divide the data set randomly into two parts;
 a training set of about 2/3 of the pixels used for
estimation  and learning (training set) and a test set of about 1/3
of the pixels used for evaluation of the prediction capability of
the estimated discrimination. The split was done 50 times in such a
way that the proportion  of clear and cloudy pixels of the whole
original data set was respected. The results are summarized as
boxplots in the following figure.

Figure \ref{fig1classificationlandwater} shows the boxplots of
misclassification errors
for the
various classification methods over 50 random splits for land (left)
and sea (right). For the land pixels, apart the NPDA method which
has a poor behavior, none of the other three methods clearly stands
out and they all perform essentially well. For the sea panels (cf.
the right panel of Figure~\ref{fig1classificationlandwater}) we get
different conclusions. Here the boxplots clearly indicate that our
noisy IFA classification method has the smallest error.
Finally, Figure \ref{fig2classificationlandwater} shows
the cloud mask overimposed to the analyzed area.

\section{Conclusions}

We have considered  multivariate density estimation with
dimensionality reduction expressed in terms of noisy independent
factor analysis (IFA) model. In this model the data are generated by
a (small) number of latent independent components having unknown
non-Gaussian distributions and observed in Gaussian noise.

Without assuming that either the number of components or the mixing
matrix are known, we have shown that the densities of this form can
be estimated with a fast rate. Using the mirror averaging
aggregation algorithm, we constructed a density estimator which
achieves a nearly parametric rate $\log^{1/4}{n}/\sqrt{n}$,
independent of the dimension of the data.

We then applied these density estimates to construct nonparametric
plug-in classifiers and have shown that they achieve, within a
logarithmic factor independent of $d$, the best obtainable rate of
the excess Bayes risk.

 These theoretical results were supported by numerical
simulations and by an application to a complex data set from a
remote sensing experiment in which our IFA classifier outperformed
several commonly used classification methods.
Implementation of the
IFA-based density estimator and of the related classifier is
computationally intensive; therefore an efficient
computational algorithm has been developed that makes mirror averaging
aggregation feasible from computational point of view. 

\renewcommand{\theequation}{A.\arabic{equation}}
\setcounter{equation}{0}  
\section*{APPENDIX: PROOFS}  

\noindent {\bf Proof of (\ref{rate1})}. Note that (\ref{conv1})
implies that the Fourier transform  $\varphi_{\bf
X}(\mathbf{u})=\int_{\RR^d}p_{\mathbf{X}}(\mathbf{x})e^{i\mathbf{x}^T\mathbf{u}}d\mathbf{x}$
of the density $p_{\mathbf{X}}$ satisfies the inequality
\begin{equation} \label{chf1}
|\varphi_{\bf X}(\mathbf{u})| \le e^{-\sigma^2 \|\mathbf{u}\|^2/2}
\end{equation}
for all $\mathbf{u} \in \RR^d$, where $\|\cdot\|$ denotes the
Euclidean norm in $\RR^d$. Define the kernel estimator
\begin{equation*}
\hat{p}^*_n(\mathbf{x})=\frac{1}{nh^d}\sum_{i=1}^n
K\left(\frac{\mathbf{X}_i-\mathbf{x}}{h}\right)
\end{equation*}
with the kernel $K: \RR^d \rightarrow \RR$, such that
$K(\mathbf{x})=\prod_{k=1}^d K_0(x_k)$,
$\mathbf{x}^T=(x_1,x_2,...,x_d)$, where $K_0$ is the sinc kernel:
$K_0(x) = \frac{\sin x}{\pi x}$, for $x \not= 0$, and $K(0)=1/\pi$,
with the Fourier transform $\Phi^{K_0}(t)=I(|t|\le 1)$.

Using Plancherel theorem and Theorem 1.4 on p.\ 21 of Tsybakov
(2009),
we have
\begin{eqnarray*}
\EE\|\hat{p}^*_n-p_{\mathbf{X}}\|^2_2&=&\frac{1}{(2\pi)^d}
E\|\varphi_n
\Phi^K-\varphi_{\bf X}\|^2_2 \\
&\le&\frac{1}{(2\pi)^d}\left[\int
|1-\Phi^K(h\mathbf{u})|^2|\varphi_{\bf
X}(\mathbf{u})|^2d\mathbf{u}+\frac{1}{n}\int
|\Phi^K(h\mathbf{u})|^2d\mathbf{u}\right] ,
\end{eqnarray*}
where $\varphi_n(\mathbf{u})=n^{-1}\sum_{j=1}^n
e^{i\mathbf{X}^T_j\mathbf{u}}$ is the empirical characteristic
function and $\Phi^K(\mathbf{v})$ is the Fourier transform of $K$.
Note that $\Phi^K(\mathbf{v})=\prod_{j=1}^d I\{|v_j|\le 1\}$ where
$v_j$ are the components of $\mathbf{v}\in\RR^d$.
 Now, for the bias term we have, using (\ref{chf1}),
\begin{eqnarray*}
\int |1-\Phi^K(h\mathbf{u})|^2|\varphi_{\bf
X}(\mathbf{u})|^2d\mathbf{u}&=& \int I\left\{\exists j: |u_j| >
\frac{1}{h}\right\}|\varphi_{\bf X}(\mathbf{u})|^2d\mathbf{u} \\
&\le&\int I\left\{\exists j: |u_j| > \frac{1}{h}\right\}e^{-\sigma^2
\mathbf{u}^2/4}e^{-\sigma^2 \mathbf{u}^2/4}d\mathbf{u} \\ &\le&
e^{-\sigma^2 /{4h^2}} \int e^{-\sigma^2
\mathbf{u}^2/4}d\mathbf{u}=e^{-\sigma^2 /{4h^2}}
\left(\frac{4\pi}{\sigma^2}\right)^{d/2}.
\end{eqnarray*}
Next, the variance term
$$\frac{1}{n}\int |\Phi^K(h\mathbf{u})|^2d\mathbf{u}=\frac{1}{n} \prod_{j=1}^d
\int I\left\{|u_j|\le \frac{1}{h}\right\}du_j=\frac{2^d}{nh^d}.$$ Combining the
last two expressions, we get
$$\EE\|\hat{p}^*_n-p_{\mathbf{X}}\|^2_2 \le C \left(e^{-\sigma^2 /{4h^2}}+\frac{1}{nh^d}\right)$$
with some constant $C>0$. Taking here $h=\sigma (4\log n)^{-1/2}$,
we get (\ref{rate1}).  \hfill $\Box$


\noindent
{\bf Proof of Proposition \ref{prop1}}.
W.l.o.g. we will
suppose here that ${\bf a}_k$ are the canonical basis vectors in
$\RR^d$. Note first that the proof of (\ref{rate1}) with $d=1$
implies that the estimators ~(\ref{kern1}) achieve the convergence
rate of $(\log n)^{1/2}/n$ for the quadratic risk:
\begin{equation}\label{rate2}
\EE\|\hat{g}_{k}-g_{k}\|_2^2=\mathcal{O}( (\log n)^{1/2} / n)  \quad
\forall k=1,\dots,m.
\end{equation}
Denoting $C>0$ a constant, not always the same, we have for the
estimator (\ref{convest})
\begin{eqnarray*}
\EE \| \hat{p}_{n,m,A} - p_{\mathbf{X}}\|_2^2 &\le& C \EE
\left\|\prod_{j=1}^m \hat{g}_j-\prod_{j=1}^m g_j\right\|_2^2 = C\EE
\left[\left\|\sum_{k=1}^m \prod_{j=1}^{k-1}
g_j(\hat{g}_k-g_k)\prod_{j=k+1}^{m}
\hat{g}_j\right\|_2^2\right] \\
&\le& C \sum_{k=1}^m \EE \left[\left\|\prod_{j=1}^{k-1}g_j\right\|_2^2
\|\hat{g}_k-g_k\|_2^2 \left\|\prod_{j=k+1}^{m} \hat{g}_j\right\|_2^2\right]\\
&\le& C \sum_{k=1}^m  \prod_{j=1}^{k-1}\|g_j\|_2^2 \EE
\left[\|\hat{g}_k-g_k\|_2^2 \prod_{j=k+1}^{m} \|\hat{g}_j\|_2^2\right]\\
&\le& C \max_{k=1}^m  \EE \big[\|\hat{g}_k-g_k\|_2^2
\prod_{j=k+1}^{m} \|\hat{g}_j\|_2^2\big],
\end{eqnarray*}
where $\prod_{i=l}^ua_i=1$ when $l>u$ and we have used that the
$L_2$-norms of $g_j$  are bounded for all $j=1,\dots,m$. The latter
is due to the fact that, by Young's inequality (see, e.g., Besov et
al., 1979), $\|g_j\|_2\le \|\phi_{1,\sigma^2}\|_2 \int
p_{S_j}=\|\phi_{1,\sigma^2}\|_2$.

We now evaluate the $L_2$-norms of $\hat{g}_j$. By separating the
diagonal and off-diagonal terms,
\begin{eqnarray}\label{kkk}
\|\hat{g}_j\|_2^2 = \frac1{nh}\int K_0^2 + \frac1{n^2} \sum_{i\ne
m}\frac1{h}K^*\left(\frac{Y_i-Y_m}{h}\right),
\end{eqnarray}
with the convolution kernel $K^*=K_0*K_0$ and we write for brevity
$Y_i={\bf a}_j^TX_i$. The second term in (\ref{kkk}) is a
$U$-statistic that we will further denote by $U_n$. Since all the
summands $\frac1{h}K^*\left(\frac{Y_i-Y_m}{h}\right)$ in $U_n$ are
uniformly $\le C/h$, by Hoeffding inequality for $U$-statistics
(Hoeffding 1963)
we get
\begin{eqnarray}\label{hh}
P(|U_n- E(U_n)| >t) \le 2 \exp(- cnh^2t^2)
\end{eqnarray}
for some constant $c>0$ independent of $n$. On the other hand, it is
straightforward to see that there exists a constant $C_0$ such that
$|E(U_n)|\le C_0$. This and (\ref{hh}) imply:
\begin{eqnarray}\label{hh1}
P(|U_n| >2C_0) \le 2 \exp(- c'nh^2)
\end{eqnarray}
for some constant $c'>0$ independent of $n$. From (\ref{kkk}) and
(\ref{hh1}) we get
\begin{eqnarray}\label{hh2}
P({\cal A}) \le 2d \exp(- c'nh^2),
\end{eqnarray}
for the random event ${\cal A}= \{\exists j: \, \|\hat{g}_j\|_2^2
\ge C_1\}$, where $C_1= 2C_0+\int K_0^2/(nh)$.

Using (\ref{hh2}), (\ref{rate2}) and the fact that $\|g_j\|_2^2$ and
$\|\hat{g}_j\|_2^2$ are uniformly $\le C/h$ we find
\begin{eqnarray*}
\EE \left[\|\hat{g}_k-g_k\|_2^2 \prod_{j=k+1}^{m}
\|\hat{g}_j\|_2^2\right] &\le & \EE \left[\|\hat{g}_k-g_k\|_2^2
\prod_{j=k+1}^{m} \|\hat{g}_j\|_2^2 I\{{\cal A}\}\right] \\
&&+  \ (C_1)^{m-k} \EE \left[\|\hat{g}_k-g_k\|_2^2 I\{{\cal A}^c\}\right]\\
&\le & (C/h)^{m-k+1}P\{{\cal
A}\} + C(\log n)^{1/2}/n\\
&\le & C h^{-(m-k+1)}\exp(- c'nh^2) + C(\log n)^{1/2}/n\\
&\le & C(\log n)^{1/2}/n.
\end{eqnarray*}
Thus, the proposition follows.  \hfill $\Box$


\noindent
{\bf Proof of (\ref{bdd1})}.
We will show first that for
some constant $C>0$ and for all $j=1,...,M$
\begin{equation}\label{sup1}
\PP(\|\hat{g}_{j}\|_{\infty,[-1,1]}>C) \le \frac{1}{n^{1/2}h^{3/2}},
\end{equation}
where $\|f\|_{\infty,[-1,1]}=\sup_{t\in [-1,1]} |f(t)|$ for
$f:\RR\to \RR$.
 Note that the sinc kernel $K_0$ satisfies the inequality
$|K_0(u)| \le 1/{\pi}$ for all $u\in \RR$. Now because
$$\|\hat{g}_{j}\|_{\infty,[-1,1]} \le \EE\|\hat{g}_{j}\|_{\infty,[-1,1]} +
\|\hat{g}_{j}-\EE\hat{g}_{j}\|_{\infty,[-1,1]}$$ and
$$\left|\EE\hat{g}_{j}(t)\right|=\left|\int K_0(u)g_j(t-uh)du\right| \le \frac{1}{\pi},
\quad \forall t \in \RR,$$  we have
\begin{equation}\label{sup2}
\PP(\|\hat{g}_{j}\|_{\infty,[-1,1]}>C) \le
\PP\left(\|\hat{g}_{j}-\EE\hat{g}_{j}\|_{\infty,[-1,1]} >
C-\frac{1}{\pi}\right).
\end{equation}
Now for $\eta(t):=\hat{g}_{j}(t)-\EE\hat{g}_{j}(t)$ we have
\begin{equation}
\label{cond1}
\begin{split}
\EE(\eta(t+\Delta)-\eta(t))^2 =&
\frac{1}{nh^2}\mathrm{Var}\left(K_0\left(\frac{t+\Delta-Z}{h}\right)-K_0\left(\frac{t-Z}{h}\right)\right)\\
\le&\frac{1}{nh^2}\int
\left(K_0\left(\frac{t+\Delta-z}{h}\right)-K_0\left(\frac{t-z}{h}\right)\right)^2g_k(z)dz\\
 \le&
\frac{C_0^2}{nh^3} \Delta^2
\end{split}
\end{equation}
for $t, \Delta \in [-1,1]$, where we used that $|K'_0(u)|\le C_0$
with some constant $C_0$ for all $u\in \RR$. Also, the standard
bound for the variance of kernel estimator $\hat{g}_{j}$ gives
\begin{equation}\label{cond2}
\EE\eta^2(t) \le \frac{C_2}{nh}, \quad \forall t \in [-1,1]
\end{equation}
 with
$C_2=\int K^2_0(u)du$. Now (\ref{cond1}) and (\ref{cond2}) verify
conditions of the following lemma.

\begin{lemma} (Ibragimov and Has'minskii 1982, Appendix 1)
\label{lih} Let $\eta(t)$ be a continuous real-valued random
function defined on $\RR^d$ such that, for some $0<H<\infty$ and
$d<a<\infty$ we have
\begin{eqnarray*}
&&\EE|\eta(t+\Delta) - \eta(t)|^a \le H \|\Delta\|^a, \qquad
\forall \ t,\Delta \in \RR^d, \\
&&\EE|\eta(t)|^a \le H, \qquad \forall \ t\in \RR^d.
\end{eqnarray*}
Then for every $\delta>0$ and $t_0\in \RR^d$ such that $\|t_0\|\le
D$,
\begin{equation*}
\EE \left[ \sup_{t: \|t-t_0\|\le \delta} |\eta(t) -
\eta(t_0)|\right] \le B_0(D+\delta)^d H^{1/a}\delta^{1-d/a}
\end{equation*}
where $B_0$ is a finite constant depending only on $a$ and $d$.
\end{lemma}

Applying this lemma with  $d=1$, $a=2$, $H=\frac{C_0^2}{nh^3}$,
$t_0=0$, and $\delta=1$, we get
$$
\EE\sup_{t \in[-1,1]} |\eta(t)| \le \EE\sup_{t \in[-1,1]}
|\eta(t)-\eta(0)| +\EE |\eta(0)| \le
\frac{C_3}{n^{1/2}h^{3/2}}+\frac{C^{1/2}_2}{(nh)^{1/2}}\le
\frac{C_4}{n^{1/2}h^{3/2}}.
$$
Applying now in (\ref{sup2}) Markov inequality and choosing
$C=C_4+1/\pi$, we obtain (\ref{sup1}).

Next, assume w.l.o.g. that $B$ is the unit ball in $\RR^d$. We note
that (\ref{sup1}) implies
\begin{eqnarray*}
\PP\left(\left\|\prod_{j=1}^k \hat{g}_j\right\|_{\infty,B} > C^k\right)&\le&
\PP\left(\prod_{j=1}^k \|\hat{g}_j\|_{\infty,[-1,1]} > C^k\right) \\
&\le&\PP(\cup_{j=1}^k\{\|\hat{g}_j\|_{\infty,[-1,1]} > C\})\le
\frac{k}{n^{1/2}h^{3/2}}.
\end{eqnarray*}

Using this and definition (\ref{estk}) of $\hat{p}_k$ we have that
\begin{eqnarray*}
\EE \|\hat{p}_k\|_{\infty,B} &\le& (2\pi \sigma^2)^{(d-k)/2}\EE
\left\|\prod_{j=1}^k \hat{g}_j\right\|_{\infty,B} \\
&\le&(2\pi \sigma^2)^{(d-k)/2}\left[C^k+ \EE \left\|\prod_{j=1}^k
\hat{g}_j\right\|_{\infty,B}I\left\{\left\|\prod_{j=1}^k
\hat{g}_j\right\|_{\infty,B}> C^k\right\}\right]\\
&\le&(2\pi\sigma^2)^{(d-k)/2}\left[C^k+\frac{1}{(\pi
h)^k}\frac{k}{n^{1/2}h^{3/2}}\right],
\end{eqnarray*}
where we also used the fact that $\|\hat{g}_j\|_{\infty,[-1,1]}\le
(\pi h)^{-1}$ for all $j=1,...,k$. Since $h \asymp (\log n)^{-1/2}$,
we get that, for some constant $L_k$,
$$
\EE \|\hat{p}_k\|_{\infty,B} \le L_k, \quad \forall k=1,...,M,
$$
and (\ref{bdd1}) follows with $L'=\max (L_1,L_2,...,L_M)$.  \hfill $\Box$

\noindent {\bf Proof of Theorem \ref{th1}}.
To prove the theorem we use Corollary 5.7 in
Juditsky, Rigollet and Tsybakov~(2008), which implies that for
$\beta = 12 {\hat L}$ the corresponding aggregate estimator
$\tilde{p}_n$ satisfies:
\begin{equation}\label{ineq1}
\EE_{{\cal D}_2}  \| \tilde{p}_n - p_\mathbf{X}\|_{2}^2  \leq
\min_{k=1,\dots,M} \| \hat{p}_{n_1,k,{\hat B}_k}-
p_\mathbf{X}\|_{2}^2 + \frac{\beta \log M}{n_2},
\end{equation}
where $\EE_{D_2}$ denotes the expectation over the second,
aggregating subsample. Here $\hat{p}_{n_1,k,{\hat B}_k} $ are the
estimators constructed from the first, training subsample ${\cal
D}_1$, which is supposed to be frozen when applying the result of
Juditsky, Rigollet and Tsybakov~(2008) and the inequality holds for
any fixed training subsample. Taking expectation in
inequality~(\ref{ineq1}) with respect to the training subsample,
using that, by construction, $\tilde{p}_n$ and $\hat{p}_{n_1,k,{\hat
B}_k}$ vanish outside $B$, and interchanging the expectation and the
minimum on the right hand side we get
\begin{equation*}
\EE  \| \tilde{p}_n - p_\mathbf{X}\|_{2,B}^2  \leq
\min_{k=1,\dots,M} \EE \| \hat{p}_{n_1,k,{\hat B}_k}-
p_\mathbf{X}\|_{2,B}^2 + \frac{\log M} {n_2}\EE \beta,
\end{equation*}
where now $\EE$ is the expectation over the entire sample.

Recalling now that $M < d$,  $n_2 = [ {cn}/{\sqrt{\log n}}]$, and
that $\EE \beta \le C$ by (\ref{bdd1}), we obtain
\begin{equation}
\label{ineq12} \EE  \| \tilde{p}_n - p_\mathbf{X}\|_{2,B}^2  \leq
\min_{k=1,\dots,M} \EE \| \hat{p}_{n_1,k,{\hat B}_k}-
p_\mathbf{X}\|_{2,B}^2 + \frac{C (\log n)^{1/2}}{n}.
\end{equation}
Now,
\begin{equation}
\label{ineq13} \min_{k=1,\dots,M} \EE \| \hat{p}_{n_1,k,{\hat B}_k}-
p_\mathbf{X}\|_{2,B}^2 \leq \EE \| \hat{p}_{m,\hat{A}}-
p_\mathbf{X}\|_{2,B}^2,
\end{equation}
where $\hat{A}={\hat B}_m$ is the estimate of $A$ with the true rank
$m$ and we set for brevity $\hat{p}_{m,A}\equiv\hat{p}_{n_1,m,A}$.
Since $p_\mathbf{X}=p_{m,A}$, we have
\begin{equation}\label{ineq2}
\|\hat{p}_{m, \hat{A}} - p_\mathbf{X} \|_{2,B}^2 \leq 2 ( \|
\hat{p}_{m,\hat{A}} - \hat{p}_{m,A}\|_{2,B}^2 + \| \hat{p}_{m,A} -
p_{m,A}\|_{2,B}^2).
\end{equation}
Since $n_1= n(1+o(1))$, by Proposition \ref{prop1} we get
\begin{equation}\label{ineq21}
 \EE \| \hat{p}_{m,A} - p_{m,A}\|_{2,B}^2 = \mathcal{O}( (\log n)^{1/2} / n).
\end{equation}
It remains to prove that
\begin{equation}\label{ineq3}
\EE  \| \hat{p}_{m,\hat{A}} - \hat{p}_{m,A}\|_{2,B}^2 = \mathcal{O}(
(\log n)^{1/2} / n).
\end{equation}
Denoting $G_{\mathbf{x}}(A)=\left( \frac{1}{2\pi
\sigma^2}\right)^{(d-m)/2} \exp\left\{ - \frac{1}{2\sigma^2}
\mathbf{x}^T (\mathbf{I}_d - A A^T) \mathbf{x} \right\}$ and by
$\mathbf{\hat{a}}_{j}$ and $\mathbf{a}_{j}$ the columns of $\hat{A}$
and $A$, respectively, we can write (see (\ref{convest}) and
(\ref{estk})),
\begin{eqnarray*}
\| \hat{p}_{m,\hat{A}} - \hat{p}_{m,A}\|_{2,B}=
\|G_{\mathbf{x}}(\hat{A})\prod_{j=1}^m \hat{g}_{j}(
\mathbf{\hat{a}}_{j}^T \mathbf{x})-G_{\mathbf{x}}(A)\prod_{j=1}^m
\hat{g}_{j}( \mathbf{a}_{j}^T \mathbf{x})\|_{2,B}\\
\le C \|\prod_{j=1}^m \hat{g}_{j}( \mathbf{\hat{a}}_{j}^T
\mathbf{x})-\prod_{j=1}^m g_{j}( \mathbf{\hat{a}}_{j}^T
\mathbf{x})\|_{2,B}+C \|\prod_{j=1}^m \hat{g}_{j}( \mathbf{a}_{j}^T
\mathbf{x})-\prod_{j=1}^m g_{j}( \mathbf{a}_{j}^T
\mathbf{x})\|_{2,B}+ \\
\|G_{\mathbf{x}}(\hat{A})\prod_{j=1}^m g_{j}( \mathbf{\hat{a}}_{j}^T
\mathbf{x})-G_{\mathbf{x}}(A)\prod_{j=1}^m g_{j}( \mathbf{a}_{j}^T
\mathbf{x})\|_{2,B}=:I_1+I_2+I_3.
\end{eqnarray*}
As in the proof of Proposition 1 we get $\EE I_i^2 =\mathcal{O}(
(\log n)^{1/2} / n)$, $i=1,2$. Next, we show that $\EE I_3^2
=\mathcal{O}(1/ n)$. We write $I_3 \le I_{3,1}+I_{3,2}$ where
\begin{eqnarray*}
I_{3,1} = \|G_{\mathbf{x}}(\hat{A})-G_{\mathbf{x}}(A)\|_{2,B}
\|\prod_{j=1}^m g_{j}( \mathbf{a}_{j}^T \mathbf{x})\|_{2,B},\\
I_{3,2} = C \|\prod_{j=1}^m g_{j}( \mathbf{\hat{a}}_{j}^T
\mathbf{x})-\prod_{j=1}^m g_{j}( \mathbf{a}_{j}^T
\mathbf{x})\|_{2,B}.
\end{eqnarray*}
To bound these terms we will systematically use the fact that
$\|\prod_{j=k}^l g_{j}( \mathbf{a}_{j}^T \mathbf{x})\|_{2,B}\le C$
for all $1\le k\le l \le m$ (and the same with
$\mathbf{\hat{a}}_{j}$ instead of $\mathbf{a}_{j}$). This fact, the
definition of $G_{\mathbf{x}}(\cdot)$ and the boundedness of the
Frobenius norms of $A$ and $\hat{A}$ imply that $I_{3,1} \le
C\|A-\hat{A}\|_{F}$, where $\|M\|_{F}$ denotes the Frobenius norm of
matrix $M$. Now, $\EE \|\hat{A}-A\|_{F}^2=\mathcal{O}(1/n)$, which
follows from Lemma A.1 of Kneip and Utikal (2001) and the assumed
moment condition on ${\bf X}$. Thus, $\EE I_{3,1}^2
=\mathcal{O}(1/n)$.  We also get $\EE I_{3,2}^2 =\mathcal{O}(1/n)$.
This follows from the Lipschitz continuity of $g_j(\cdot)$ and from
the fact that (cf. proof of Proposition 1):
\begin{eqnarray*}
\EE I_{3,2}^2 \le C \sum_{k=1}^m \EE\left[ \Big\|\prod_{j=1}^{k-1}
g_{j}( \mathbf{a}_{j}^T \mathbf{x})\Big\|_{2,B}^2 \, \| g_{k}(
\mathbf{a}_{k}^T \mathbf{x}) - g_{k}( \mathbf{\hat{a}}_{k}^T
\mathbf{x})\|_{2,B}^2 \,
  \Big\|\prod_{j=k+1}^{m} g_{j}( \mathbf{\hat{a}}_{j}^T
\mathbf{x})\Big\|_{2,B}^2\right]
\end{eqnarray*}
So, we have $\EE I_{3}^2 =\mathcal{O}(1/n)$. This finishes the proof
of (\ref{ineq3}).

Inequalities (\ref{ineq2}), (\ref{ineq21}), and (\ref{ineq3}) give
$$\EE \| \hat{p}_{m,\hat{A}}- p_\mathbf{X}\|_{2,B}^2  \leq
\mathcal{O}((\log n)^{1/2}/{n}),$$
 which together with (\ref{ineq12}) and
(\ref{ineq13}) completes the proof. \hfill $\Box$

\noindent {\bf Proof of Proposition \ref{prop2}}. For any classifier
$T$ we have
\begin{eqnarray*}
R_B(T) - R_B(T^{*}) &=& \sum_{j=1}^J \pi_j \int_{B} (I(
T(\mathbf{x}) \not=j) - I( T^*(\mathbf{x}) \not=j))f_j(\mathbf{x}) d
\mathbf{x}\\
&=& \sum_{j=1}^J \pi_j \int_{B} (I( T^*(\mathbf{x}) =j) - I(
T(\mathbf{x}) =j))f_j(\mathbf{x}) d
\mathbf{x}\\
\\
&=& \int_{B} (\pi_{T^*(\mathbf{x})}f_{T^*(\mathbf{x})}(\mathbf{x}) -
\pi_{T(\mathbf{x})}f_{T(\mathbf{x})}(\mathbf{x})) d \mathbf{x}
\end{eqnarray*}
Therefore, the excess risk of the plug-in classifier $\hat{T}$ can
be written in the form
\begin{eqnarray}\nonumber
{\cal E}(\hat{T}) &\equiv& \EE (R_B(\hat{T})) - R_B(T^{*}) \\&=& \EE
\int_{B} (\pi_{T^*}f_{T^*}(\mathbf{x})-
\pi_{\hat{T}}\hat{f}_{\hat{T}}(\mathbf{x})+
\pi_{\hat{T}}\hat{f}_{\hat{T}}(\mathbf{x})-
\pi_{\hat{T}}f_{\hat{T}}(\mathbf{x})) d \mathbf{x}\label{plug1}
\end{eqnarray}
where we omit for brevity the argument $\mathbf{x}$ of $T^*$ and
$\hat{T}$. Note that, by the definition of $\hat{T}$, for all
$\mathbf{x}\in \RR^d$ we have:
\begin{eqnarray*}
\pi_{T^*}f_{T^*}(\mathbf{x})-
\pi_{\hat{T}}\hat{f}_{\hat{T}}(\mathbf{x})+
\pi_{\hat{T}}\hat{f}_{\hat{T}}(\mathbf{x})-
\pi_{\hat{T}}f_{\hat{T}}(\mathbf{x}) &\le&
\pi_{T^*}f_{T^*}(\mathbf{x})- \pi_{T^*}\hat{f}_{T^*}(\mathbf{x}) +
\pi_{\hat{T}}|\hat{f}_{\hat{T}}(\mathbf{x})-{f}_{\hat{T}}(\mathbf{x})|\\
&\le &\sum_{j=1}^J \pi_j |\hat{f}_j(\mathbf{x})-{f}_j(\mathbf{x})|.
\end{eqnarray*}
Combining the last display with (\ref{plug1}) proves the
proposition. \hfill $\Box$

\section*{References}

\begin{description}
\footnotesize{
\item
Amato, U., Antoniadis, A., and Gr\'egoire, G.\ (2003), ``Independent Component
Discriminant Analysis,'' {\it Internationl Journal of Mathematics}, 3, 735--753

\item
Anderson, T.~W., and Rubin, H.\ (1956), Statistical inference in factor analysis, in ``{\it Proceedings of the Third Berkeley Symposium on Mathematical Statistics and Probability}'' (Vol.\ V), ed.\ J.~Neyman, Berkeley and Los Angeles: University of California Press, 111--150

\item
An, Y., Hu, X., and Xu, L.\ (2006), ``A comparative investigation on
model selection in independent factor analysis,'' {\it Journal of
Mathematical Modeling and Algorithms}, 5, 447-473.

\item
Artiles, L.~M.\ (2001), ``Adaptive minimax estimation in classes of
smooth functions,'' Ph.D.\ thesis, University of Utrecht

\item
Attias, H.\ (1999), ``Independent Factor Analysis,'' {\it Neural Computation}, 11, 803--851

\item
Audibert, J.~U., and Tsybakov, A.~B.\ (2007), ``Fast learning
rates for plug-in classifiers,'' {\it The Annals of Statistics} 35, 608--633


\item
 Belitser, E., and Levit, B.\ (2001), ``Asymptotically local minimax estimation of infinitely smooth density with censored data,'' {\it Annals of the Institute of Statistical Mathematics}, 53, 289--306

\item
Besov,~O.V., Ilin,~V.P., and  Nikolskii,~S.M.\ (1979), {\it
``Integral Representations of Functions and Embedding Theorems"},
vol. 1, Wiley.

\item
Blanchard, B., Kawanabe, G.~M., Sugiyama, M., Spokoiny, V., and M\"uller, K.~R.\ (2006), ``In search of non-gaussian components
of a high-dimensional distribution,'' {\it Journal of Machine Learning
Research}, 7, 247--282.



\item
Cook, R.~D., and Li, B.\ (2002), ``Dimension reduction for conditional mean in regression,'' {\it The Annals of Statistics}, 32, 455--474

\item
Devroye, L., Gy\"orfi, L., and Lugosi, G.\ (1996), ``{\it A
Probabilistic Theory of Pattern Recognition"}, New York: Springer

\item
Fan, K. (1951), ``Maximum Properties and Inequalities for the Eigenvalues of Completely Continuous Operators,'' {\it Proc. Natl. Acad. Sci. U S A.}, 37(11), 760–-766.

\item
Glad, I.~K., Hjort, N.~L., and Ushakov, N.G. (2003), ``Correction of density estimators that are not densities,'' {\it Scandinavian Journal of Statistics}, 30, 415--427


\item
Hall, P., and Murison, R.~D.\ (1993), ``Correcting the negativity of high-order kernel density estimators,'' {\it Journal of Multivariate Analysis}, 47, 103--122

\item
Hastie, T., Tibshirani, R., and Buja, A,\ (1994), ``Flexible Discriminant Analysis by Optimal Scoring,'' {\it Journal of the American Statistical Association}, 89, 1255--1270


\item
Hoeffding, W.\ (1963), ``Probability inequalities for sums of bounded random variables,'' {\it Journal of the American Statistical Association}, 58, 13--30

\item
Hyvarinen, A., Karhunen, J., and Oja, E.\ (2001), ``{\it Independent Component Analysis}'', New York: John Wiley and Sons

\item
Ibragimov, I.~A., and Has'minskii, R.~Z. (1981), ``{\it Statistical Estimation: Asymptotic Theory},'' New York: Springer

\item
 Ibragimov, I.~A., and Khasminski\u\i, R.~Z.\ (1982), ``An estimate of the density of a distribution belonging to a class of entire functions'' (Russian),  {\it Teoriya Veroyatnostei i ee Primeneniya}, 27, 514--524


\item
Juditsky, A.~B., Nazin, A.~V, Tsybakov, A.~B., and Vayatis, N.\ (2005), ``Recursive Aggregation of Estimators by the Mirror Descent Algorithm with Averaging,'' {\it Problems of Information Transmission}, 41, 368--384

\item
Juditsky, A., Rigollet, P., and Tsybakov, A.~B.\ (2008), ``Learning by mirror averaging,'' {\it The Annals of Statistics}, 36, 2183--2206

\item
Kawanabe, M., Sugiyama, M., Blanchard, G., and M\"uller, K.~R.\ (2007), ``A new algorithm of non-Gaussian component analysis with radial kernel functions,'' {\it Annals of the Institute of Statistical Mathematics}, 59, 57--75

\item
Kneip, A., and Utikal, K.~(2001), ``Inference for density families using functional principal components analysis (with discussion),'' {\it Journal of the American Statistical Association}, 96, 519--542

\item
Montanari, A., Cal\`o, D., and Viroli, C.\ (2008), Independent factor discriminant analysis, {\it Computational Statistics and Data Analysis}, 52, 3246--3254

\item
Platnick, S., King, M.~D., Ackerman, S.~A., Menzel, W.~P, Baum, P.~A., Ridi, J.~C, and Frey, R.~A.\ (2003), ``The MODIS cloud products: Algorithms and examples from Terra,'' {\it IEEE Transactions on Geosciences and Remote Sensing}, 41, 459--473.

\item
Polzehl, J.\ (1995), ``Projection pursuit discriminant analysis,'' {\it Computational Statistics and Data Analysis}, 20, 141--157

\item
Roweis, S., and Saul, L.\ (2000), ``Nonlinear dimensionality reduction by locally linear embedding'', {\it Science}, 290, 2323--2326

\item
Samarov, A., and Tsybakov, A.~B.\ (2004), ``Nonparametric independent component analysis'', {\it Bernoulli}, 10, 565--582

\item
Samarov, A., and Tsybakov, A.~B.\ (2007), ``Aggregation of density estimators and dimension reduction,'', in {\it Advances in Statistical Modeling and Inference, Essays in Honor of K. Doksum}, Series in Biostatistics (Vol.\ 3), V. Nair (ed.), London: World Scientific, pp.\ 233--251

\item
Silverman, B.~W.\ (1982), ``Kernel density estimation using the fast Fourier transform,'' {\it Applied Statistics}, 31, 93--99

\item
Tenenbaum, J.~B., de Silva, V., and Langford, J.~C.\ (2000), ``A
global geometric framework for nonlinear dimensionality reduction,''
{\it Science}, 290, 2319--2323

\item
Tsybakov, A.~B.\ (2009), ``{\it Introduction to Nonparametric
Estimation},'' New York: Springer (2009)

\item
Wand, M.~P., and Jones, M.~C.\ (1995), ``{\it Kernel Smoothing},''
London: Chapman \& Hall/CRC

\item
Yang, Y.~(1999), ``Minimax nonparametric classification. I. Rates of
convergence. II. Model selection for adaptation.'' {\it IEEE Trans.
Inform. Theory}, 45, 2271--2292.


}

\end{description}

\newpage

\begin{table}[htb]
\begin{center}

{
\renewcommand{\baselinestretch}{1.}
\small\normalsize

\begin{tabular}{c|c}
Index & Test function \\
\hline\hline
1 & $\mathcal{G}(0,1)$ \\
2 & $\chi^2(1)$ \\
3 & $0.5\mathcal{G}(-3,1) + 0.5 \mathcal{G}(2,1)$ \\
4 & $0.4 \gamma(5) + 0.6 \gamma(13)$ \\
5 & $\chi^2(8)$ \\
6 & $t(5)$ \\
7 & $\mathrm{Double\ exponential:}\ \exp(-|x|) $ \\
\hline
\end{tabular}

}

\caption{List of basic functions considered for the numerical experiments. $\mathcal{G}(q,r)$ stands for Gaussian distribution with mean $q$ and standard deviation $r$; $\chi^2(r)$ indicates chi-square density function with $r$ degrees of freedom; $\gamma(r)$ is Gamma distribution of parameter $r$; $t(r)$ is Student distribution with $r$ degrees of freedom.}
\label{tab test}
\end{center}
\end{table}

\begin{table}[ht]
\begin{center}
{

{
\renewcommand{\baselinestretch}{1.}
\small\normalsize

\begin{tabular}{l|cc}
Experiment & Aggregated IFA & KS \\ \hline\hline
$d=2$, $n=500$ & 0.3 & 3\\
$d=3$, $n=500$ & 0.9 & 15\\
$d=5$, $n=500$ & 4 & 120\\
\hline
\end{tabular}}

}

\caption{Computational time (sec) of aggregated IFA and
KS for some test configurations.} \label{tab 5}
\end{center}
\end{table}

\newpage

\begin{figure}
\caption{Boxplot of the error criterion $I_1$ (Eq.
(\ref{index})) in the case $d=2$, Signal to Noise Ratio 3 and test
function 2 for several sample sizes.}
\label{figboxplotd2f2snr3}
\end{figure}

\begin{figure}

\caption{Boxplot of the error criterion $I_1$ (Eq.
(\ref{index})) in the case $d=3$, Signal to Noise Ratio 3 and test
function 3 for several sample sizes.}
\label{fig boxplotd3f3snr3}
\end{figure}

\begin{figure}

\caption{Boxplot of the error criterion $I_1$ (Eq. (\ref{index})) in the case $d=5$,
Signal to Noise Ratio 3 and test functions 5 and 6 for several sample sizes. }
\label{fig boxplotd5f5_6snr3}
\end{figure}

\begin{figure}
\caption{RGB image obtained from the SEVIRI sensor onboard MSG on June 30th 2006 UTC Time 11:12. }
\label{fig msg}
\end{figure}

\begin{figure}
\caption{Boxplot of the misclassifications for the considered classifiers. Results refer to land (left)
and water (right) pixels of the remote sensing data. }
\label{fig1classificationlandwater}
\end{figure}

\begin{figure}
\caption{Cloud mask estimated over a part of the region in Fig.\ \ref{fig msg} by Noisy IFA. Black: area not subject to classification; dark gray: pixels over water classified as clear; light gray: pixels over land classified as clear; white: pixels over land or sea classified as cloudy.}
\label{fig2classificationlandwater}
\end{figure}

\clearpage
\newpage

\centering
\includegraphics[clip=true, viewport=0 0 400 400]{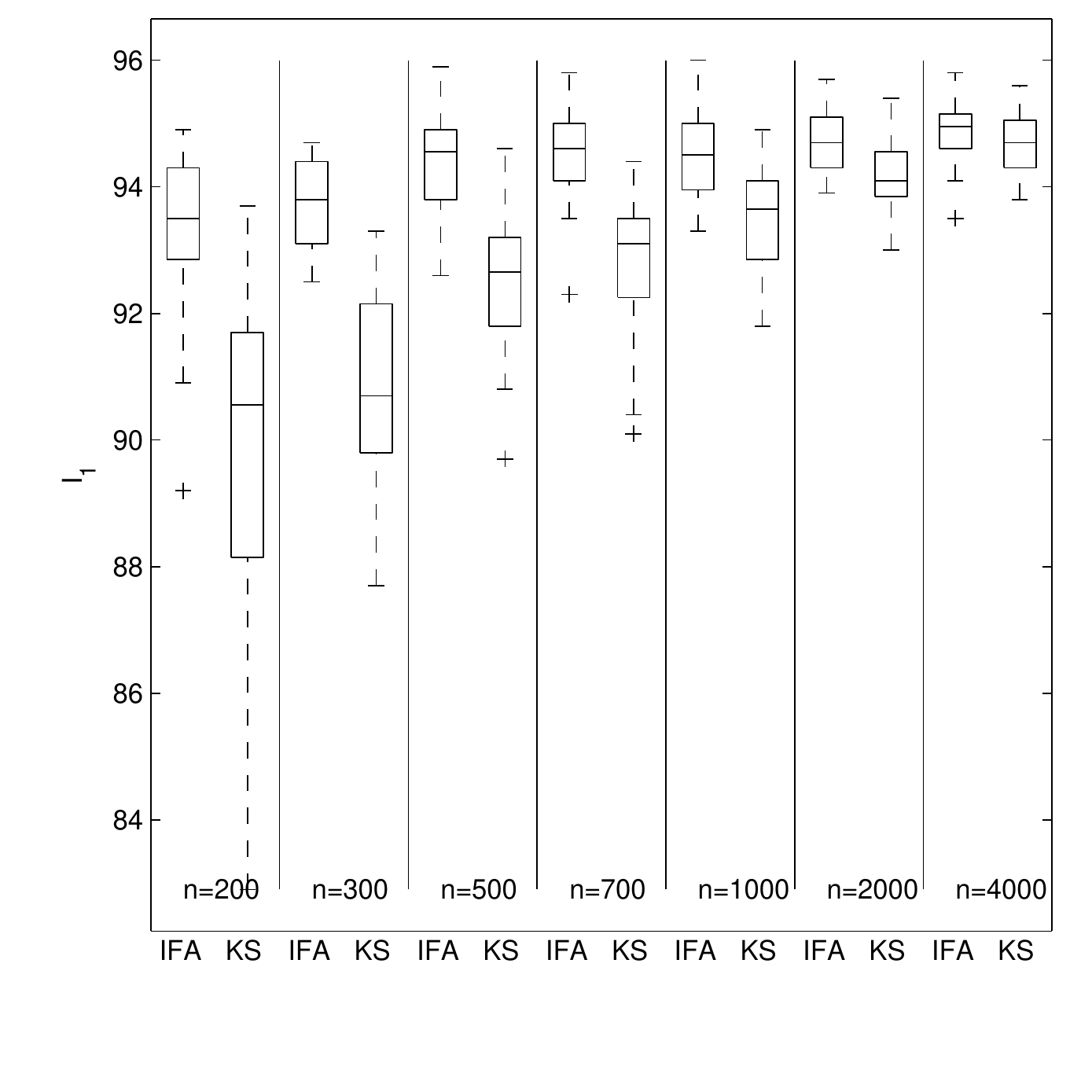}

Figure \ref{figboxplotd2f2snr3}

\includegraphics[clip=true, viewport=0 0 400 400]{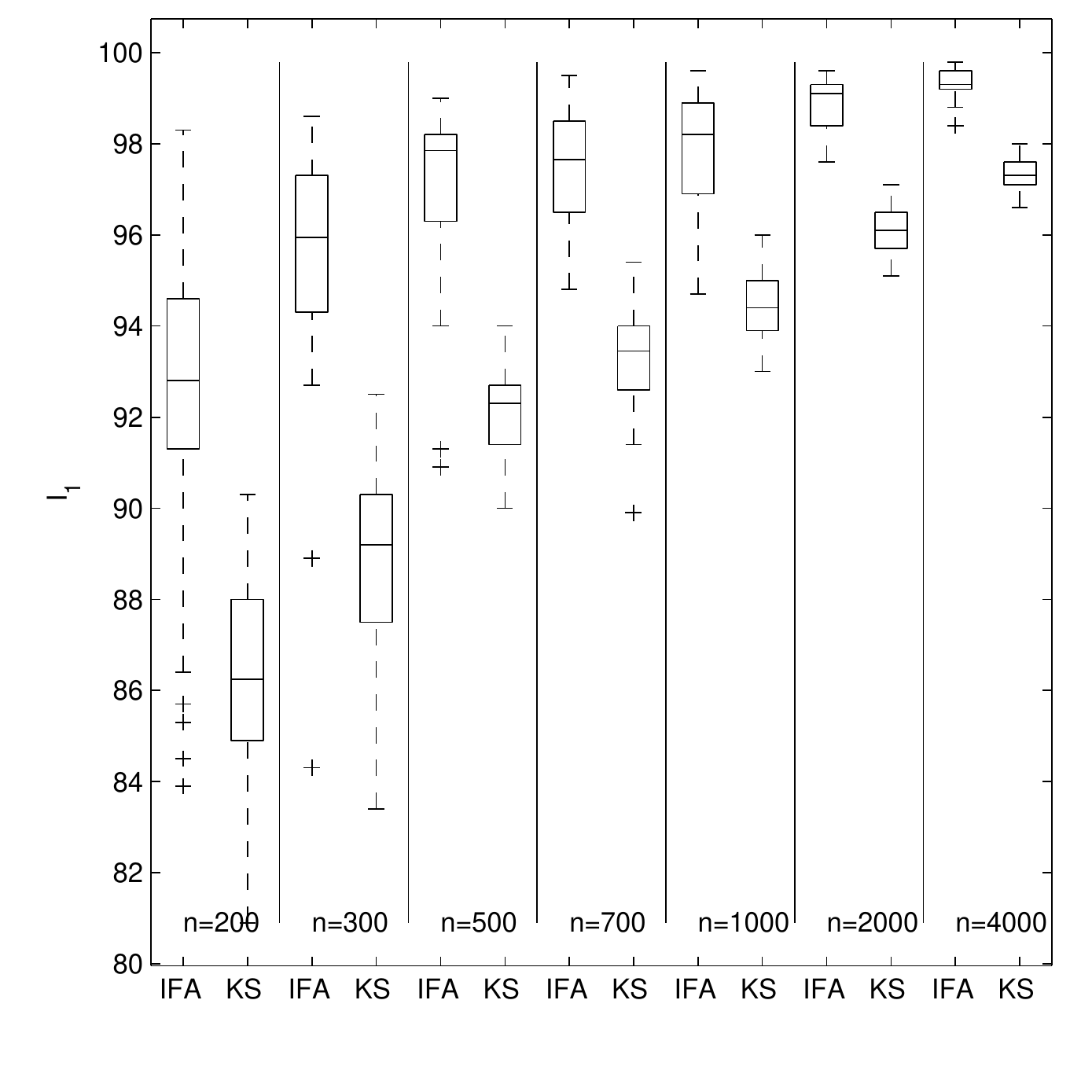}

Figure \ref{fig boxplotd3f3snr3}

\includegraphics[clip=true, viewport=0 0 400 400]{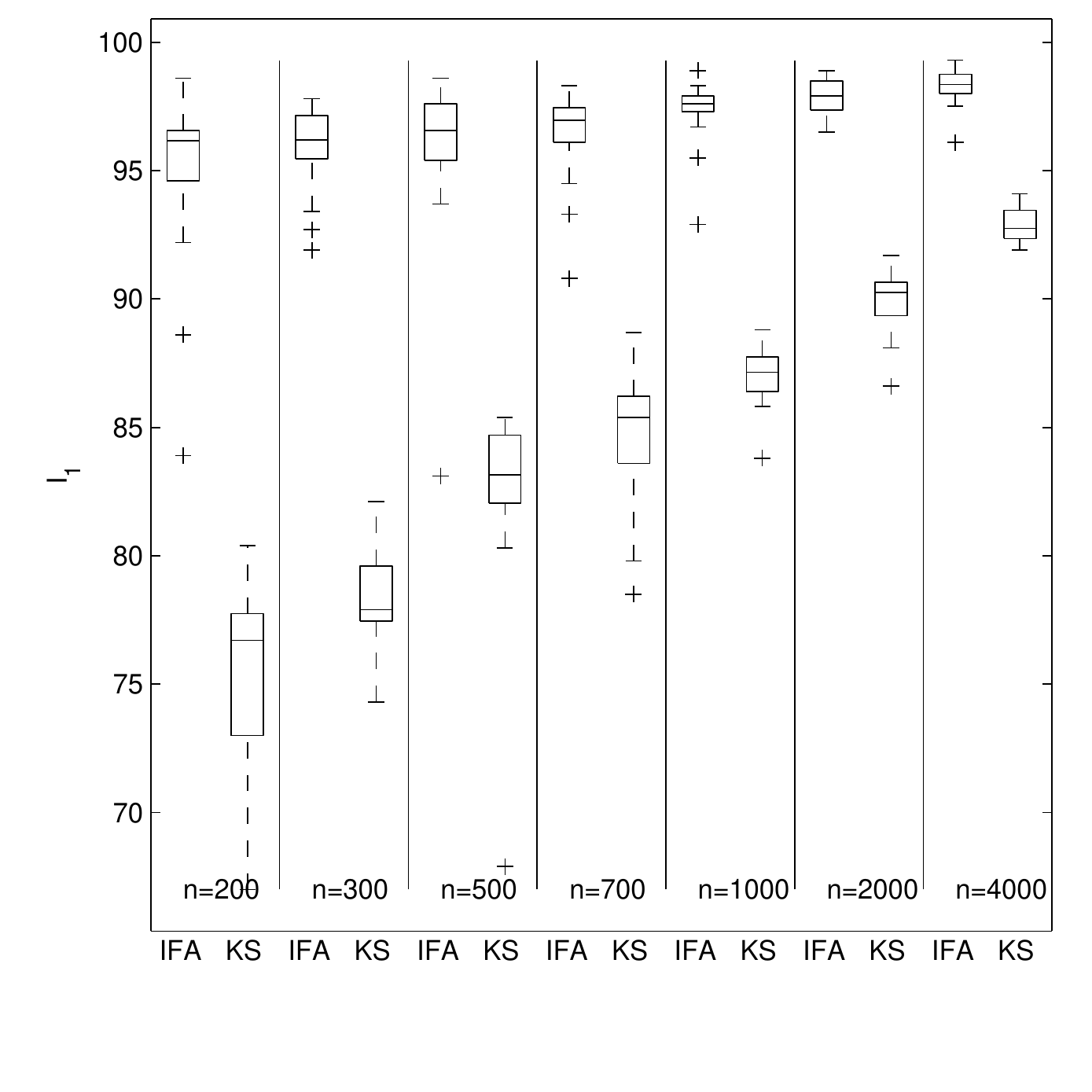}

Figure \ref{fig boxplotd5f5_6snr3}

\includegraphics[angle = 270, width=18cm, clip=true,viewport=200 200 400 700]{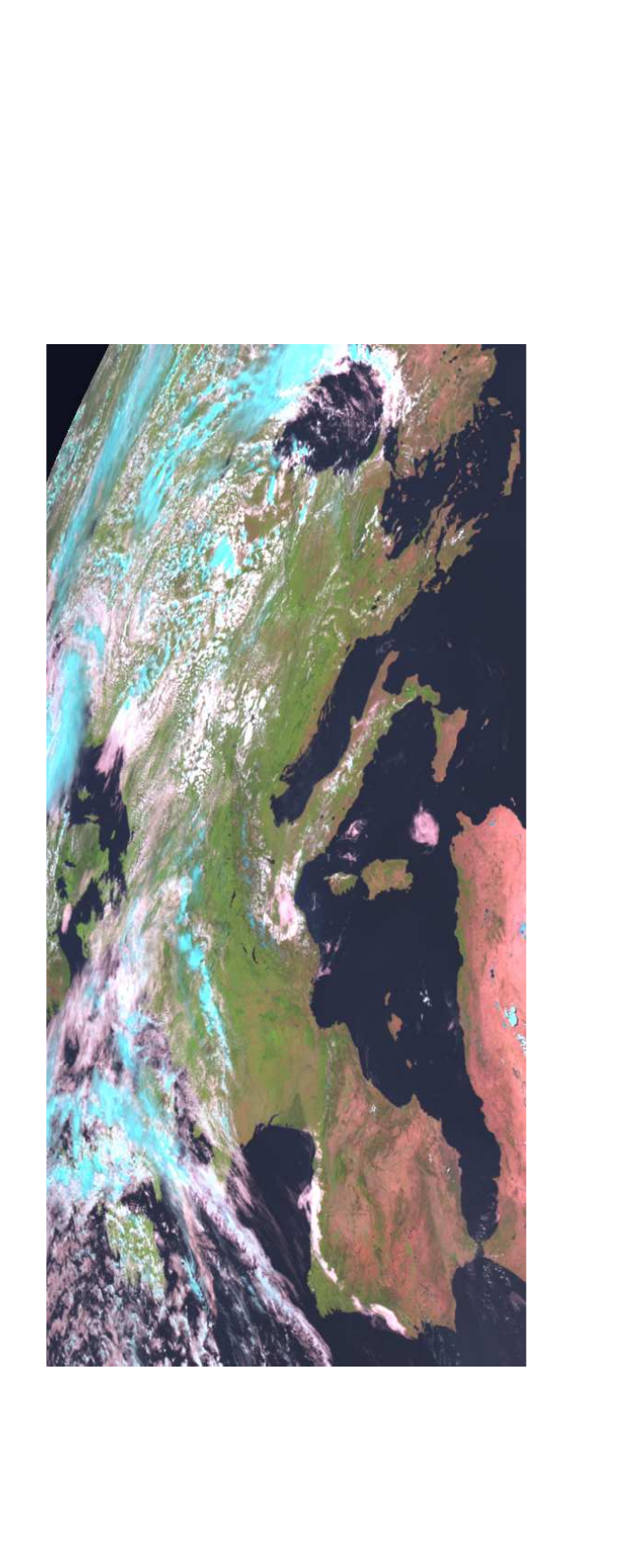}

Figure \ref{fig msg}


\includegraphics[angle = 0, width=7.5cm, clip=true,viewport=0 0 300 300]{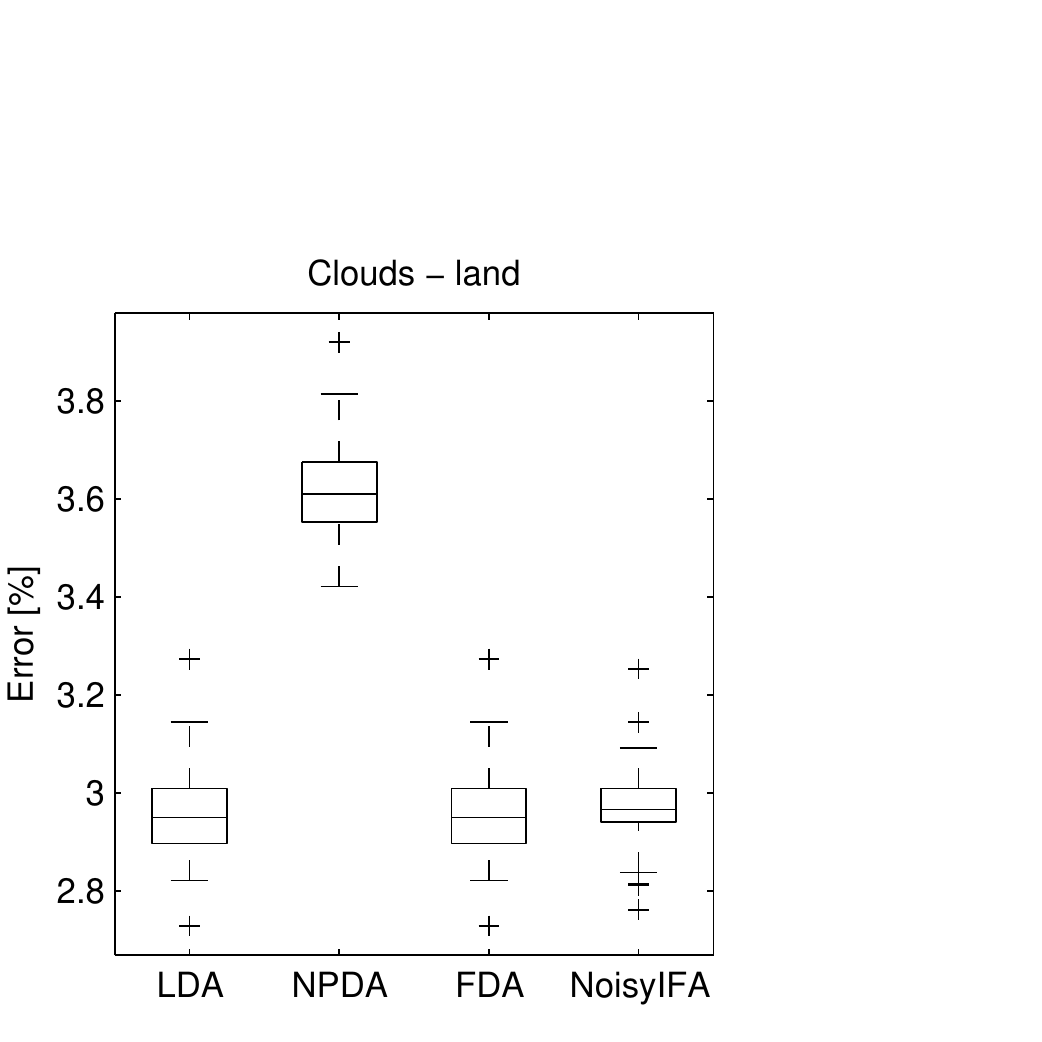}
\includegraphics[angle = 0, width=7.5cm, clip=true,viewport=0 0 300 300]{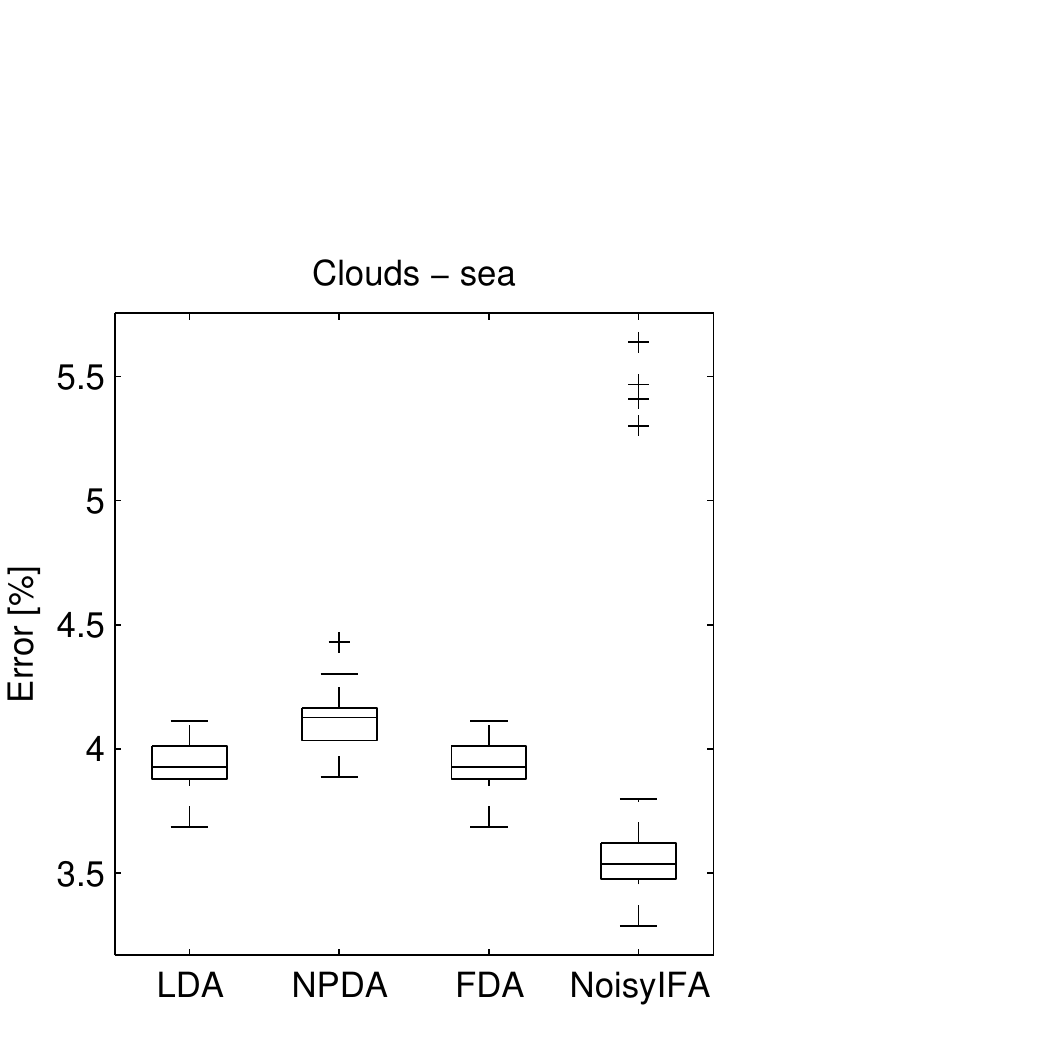}

Figure \ref{fig1classificationlandwater}

\includegraphics[angle = 0, width=18cm, clip=true,viewport=50 50 250 125]{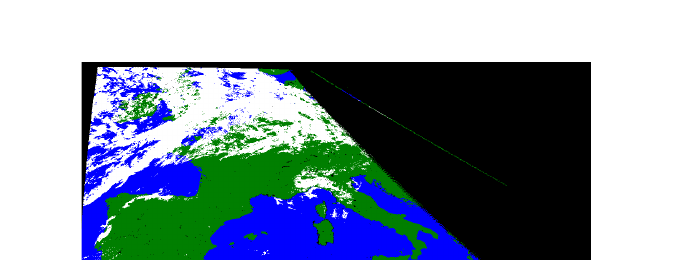}

Figure \ref{fig2classificationlandwater}

\end{document}